\documentclass[twoside,twocolumn,9pt]{article}
\usepackage{extsizes}
\usepackage[super,sort&compress,comma]{natbib} 
\usepackage[version=3]{mhchem}
\usepackage[left=1.5cm, right=1.5cm, top=1.785cm, bottom=2.0cm]{geometry}
\usepackage{balance}
\usepackage{times,mathptmx}
\usepackage{sectsty}
\usepackage{graphicx} 
\usepackage{lastpage}
\usepackage[format=plain,justification=justified,singlelinecheck=false,font={stretch=1.125,small,sf},labelfont=bf,labelsep=space]{caption}
\usepackage{float}
\usepackage{fancyhdr}
\usepackage{fnpos}
\usepackage[english]{babel}
\addto{\captionsenglish}{%
  
}
\usepackage{array}
\usepackage{droidsans}
\usepackage{charter}
\usepackage[T1]{fontenc}
\usepackage[usenames,dvipsnames]{xcolor}
\usepackage{setspace}
\usepackage[compact]{titlesec}
\usepackage{hyperref}
\usepackage{epstopdf}%This line makes .eps figures into .pdf - please comment out if not required.

\definecolor{cream}{RGB}{222,217,201}

\begin{document}

\pagestyle{fancy}
\thispagestyle{plain}
\fancypagestyle{plain}{

%%%HEADER%%%
\fancyhead[C]{\includegraphics[width=18.5cm]{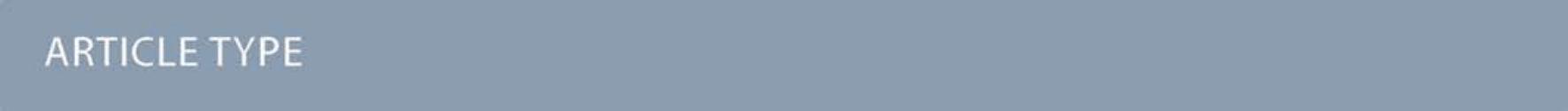}}
\fancyhead[L]{\hspace{0cm}\vspace{1.5cm}\includegraphics[height=30pt]{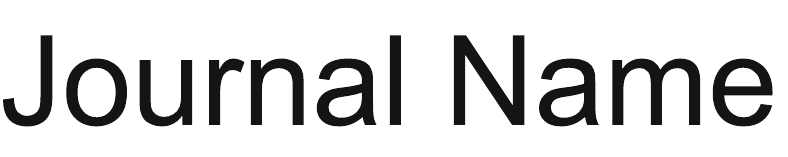}}
\fancyhead[R]{\hspace{0cm}\vspace{1.7cm}\includegraphics[height=55pt]{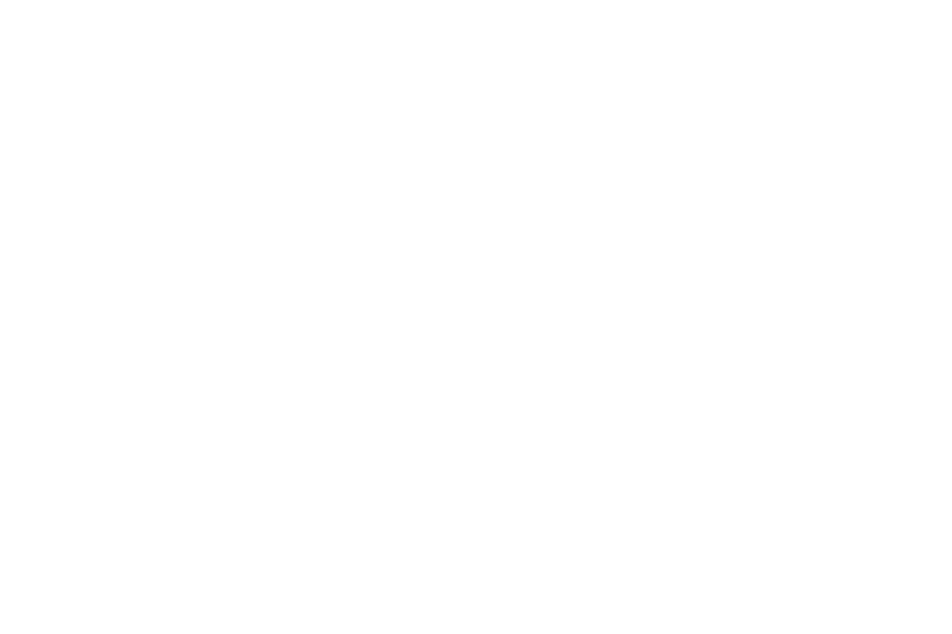}}
\renewcommand{\headrulewidth}{0pt}
}
%%%END OF HEADER%%%

%%%PAGE SETUP - Please do not change any commands within this section%%%
\makeFNbottom
\makeatletter
\renewcommand\LARGE{\@setfontsize\LARGE{15pt}{17}}
\renewcommand\Large{\@setfontsize\Large{12pt}{14}}
\renewcommand\large{\@setfontsize\large{10pt}{12}}
\renewcommand\footnotesize{\@setfontsize\footnotesize{7pt}{10}}
\makeatother

\renewcommand{\thefootnote}{\fnsymbol{footnote}}
\renewcommand\footnoterule{\vspace*{1pt}% 
\color{cream}\hrule width 3.5in height 0.4pt \color{black}\vspace*{5pt}} 
\setcounter{secnumdepth}{5}

\makeatletter 
\renewcommand\@biblabel[1]{#1}            
\renewcommand\@makefntext[1]% 
{\noindent\makebox[0pt][r]{\@thefnmark\,}#1}
\makeatother 
\renewcommand{\figurename}{\small{Fig.}~}
\sectionfont{\sffamily\Large}
\subsectionfont{\normalsize}
\subsubsectionfont{\bf}
\setstretch{1.125} %In particular, please do not alter this line.
\setlength{\skip\footins}{0.8cm}
\setlength{\footnotesep}{0.25cm}
\setlength{\jot}{10pt}
\titlespacing*{\section}{0pt}{4pt}{4pt}
\titlespacing*{\subsection}{0pt}{15pt}{1pt}
%%%END OF PAGE SETUP%%%

%%%FOOTER%%%
\fancyfoot{}
\fancyfoot[LO,RE]{\vspace{-7.1pt}\includegraphics[height=9pt]{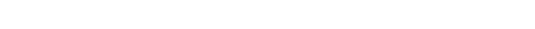}}
\fancyfoot[CO]{\vspace{-7.1pt}\hspace{13.2cm}\includegraphics{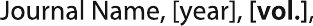}}
\fancyfoot[CE]{\vspace{-7.2pt}\hspace{-14.2cm}\includegraphics{head_foot/RF}}
\fancyfoot[RO]{\footnotesize{\sffamily{1--\pageref{LastPage} ~\textbar  \hspace{2pt}\thepage}}}
\fancyfoot[LE]{\footnotesize{\sffamily{\thepage~\textbar\hspace{3.45cm} 1--\pageref{LastPage}}}}
\fancyhead{}
\renewcommand{\headrulewidth}{0pt} 
\renewcommand{\footrulewidth}{0pt}
\setlength{\arrayrulewidth}{1pt}
\setlength{\columnsep}{6.5mm}
\setlength\bibsep{1pt}
%%%END OF FOOTER%%%

%%%FIGURE SETUP - please do not change any commands within this section%%%
\makeatletter 
\newlength{\figrulesep} 
\setlength{\figrulesep}{0.5\textfloatsep} 

\newcommand{\topfigrule}{\vspace*{-1pt}% 
\noindent{\color{cream}\rule[-\figrulesep]{\columnwidth}{1.5pt}} }

\newcommand{\botfigrule}{\vspace*{-2pt}% 
\noindent{\color{cream}\rule[\figrulesep]{\columnwidth}{1.5pt}} }

\newcommand{\dblfigrule}{\vspace*{-1pt}% 
\noindent{\color{cream}\rule[-\figrulesep]{\textwidth}{1.5pt}} }

\makeatother
%%%END OF FIGURE SETUP%%%

%%%TITLE, AUTHORS AND ABSTRACT%%%
\twocolumn[
  \begin{@twocolumnfalse}
\vspace{3cm}
\sffamily
\begin{tabular}{m{4.5cm} p{13.5cm} }

\includegraphics{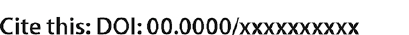} & \noindent\LARGE{\textbf{Enhanced motility in a binary mixture of active nano/microswimmers}} 
\\%Article title goes here instead of the text "This is the title"
\vspace{0.3cm} & \vspace{0.3cm} \\

%& \noindent\large{Tanwi Debnath,\textit{$^{a}$} and Pulak Kumar Ghosh\textit{$^{b\dag}$}} \\%Author names go here instead of "Full name", etc.

& \noindent\large{Debajyoti Debnath\textit{$^{1}$}, Pulak Kumar Ghosh\textit{$^{ 1}$}$^{\dag}$, Vyacheslav R. Misko\textit{$^{ 2,3}$}$^{\dag}$, Yunyun Li\textit{$^{4}$}, Fabio Marchesoni\textit{$^{4}$}, Franco Nori\textit{$^{2,6}$}} \\%Author names go here instead of "Full name", etc.

%\author{Debajyoti Debnath$^{1}$, Pulak K. Ghosh$^{1}$,  Vyacheslav R. Misko$^Vyacheslav R. Misko{2,3}$, Yunyun Li$^{4}$, Fabio Marchesoni$^{4,%5}$, and Franco Nori$^{2,6}$}

\includegraphics{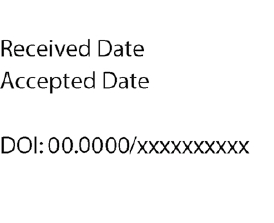} & \noindent\normalsize{It is often desirable to enhance the motility of active nano- or microscale swimmers such as, e.g., self-propelled Janus particles as agents of chemical reactions or weak sperm cells for better chances of successful fertilization.
Here we tackle this problem based on the idea that motility can be transferred from a more active guest species to a less active host species.
We performed numerical simulations of motility transfer in two typical cases, namely for interacting particles with weak inertia effect, by analyzing their velocity distributions, and for interacting overdamped particles, by studying their effusion rate.
In both cases we detected motility transfer with a motility enhancement of the host species of up to a factor of four.
This technique of motility enhancement can find applications in chemistry, biology and medicine.}

\\%The abstrast goes here instead of the text "The abstract should be..."

\end{tabular}

 \end{@twocolumnfalse} \vspace{0.6cm}

  ]
%%%END OF TITLE, AUTHORS AND ABSTRACT%%%

%%%FONT SETUP - please do not change any commands within this section
\renewcommand*\rmdefault{bch}\normalfont\upshape
\rmfamily
\section*{}
\vspace{-1cm}

%%%FOOTNOTES%%%

\footnotetext{\textit{$^{1}$Department of Chemistry, Presidency University, 86/1 College Street, Kolkata 700073, India}}

\footnotetext{\textit{$^{2}$Theoretical Quantum Physics Laboratory, RIKEN Cluster for Pioneering Research, Wako-shi, Saitama 351-0198, Japan}}

\footnotetext{\textit{$^{3}$ $\mu$Flow group, Department of Chemical Engineering, Vrije Universiteit Brussel, Pleinlaan 2, 1050 Brussels, Belgium}}

\footnotetext{\textit{$^{4}$Center for Phononics and Thermal Energy Science, School of Physics Science and Engineering, Tongji University, Shanghai 200092, People's Republic of China}}

\footnotetext{\textit{$^{5}$Istituto Nazionale di Fisica Nucleare, Sezione di Perugia, I-06123 Perugia, Italy}}

\footnotetext{\textit{$^{6}$Department of Physics, University of Michigan, Ann Arbor, Michigan 48109-1040, USA}}

\footnotetext{\textit{$^{\dag}$ Corresponding author: pulak.chem@presiuniv.ac.in; veaceslav.misco@vub.be}}

%Please use \dag to cite the ESI in the main text of the article.
%If you article does not have ESI please remove the the \dag symbol from the title and the footnotetext below.
%\footnotetext{\dag~Electronic Supplementary Information (ESI) available: [details of any supplementary information available should be included here]. See DOI: 00.0000/00000000.}
%additional addresses can be cited as above using the lower-case letters, c, d, e... If all authors are from the same address, no letter is required

%\footnotetext{\ddag~Additional footnotes to the title and authors can be included \textit{e.g.}\ `Present address:' or `These authors contributed equally to this work' as above using the symbols: \ddag, \textsection, and \P. Please place the appropriate symbol next to the author's name and include a \texttt{\textbackslash footnotetext} entry in the the correct place in the list.}

%%%END OF FOOTNOTES%%%

\section{Introduction} 
\label{Intro}
Self-propelling Janus particles (JPs), the most common class of
artificial microswimmer, have been the focus of widespread attention over the
last two decades due to their emerging applications in nano-technology and
medical sciences
\cite{review1,cataly1,pccp0,cataly2,cataly3,review2,review3}. Such particles
are made by coating one hemisphere with catalytic or photo-sensitive or
magnetic materials  \cite{cataly1,cataly2,review1}. Under appropriate
conditions, one hemisphere undergoes physical or chemical changes with
respect to the other, thus producing some local gradient in the suspension
fluid (self-phoresis). This strategy allows artificial swimmers topropel
themselves by harvesting energy from their environment.

\begin{figure}
\centering
\includegraphics[height=0.2\textwidth,width=0.5\textwidth]{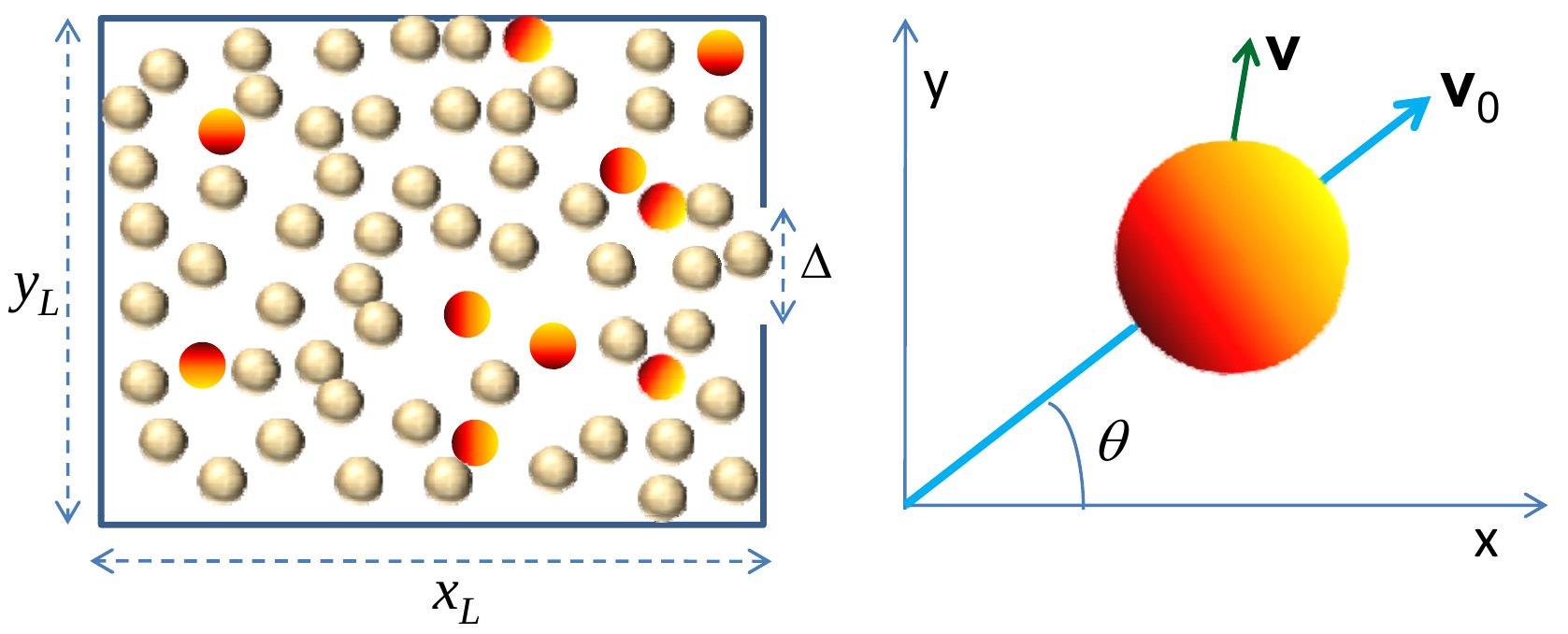}
\caption {(Color online) Left: Schematic of a mixture of self-propelled
particles in a rectangular box ($ x_L \times y_L $) with an opening of width $ \Delta $.
Right: Schematic of a two-dimensional self-propelled JP. Its dynamical, ${\vec v}$, and
self-propulsion velocities, ${\vec{v}_0}$, are depicted by distinct vectors.} \label {F1}
\end{figure}

Thanks to their self-propulsion mechanism and in contrast to their passive
peer, artificial swimmers can diffuse orders of magnitude faster
\cite{volpe1}, are capable of performing autonomous motion in periodic
structures with broken spatial symmetry  \cite{our1,nanoscale1,ai1,our2}  and
exhibit other peculiar transport properties  \cite{GNM,ourCHEMO,CHEMO1,
CHEMO2,CHEMO3}. Inspired by these unique transport features, researchers aim
to design customized JPs to be used, for instance, as ``nano-robots'' capable
of performing accurate mechanical operations
\cite{Wang,Sezer,Small16,Debnath, nanoscale2, sood}. Additional promising
technological applications have also been proposed
\cite{review1,mix1,mix2,our1,Malytska}. Among the most appealing ideas being
pursued, we mention here the recent attempt to power passive particles
through the self-propulsion mechanism of intermediary active particles
\cite{JPCM18,mix1,mix2,Lowen-mixture,Wang-mixture,our1}, to be used as
controllable stirrers. In this paper, we numerically study the velocity
distribution and effusion of active particles in a binary mixture, to
understand how to enhance motility of less active particles by adding more
active particles. Mixtures of interacting active particles (either of the
same or different kinds) behave quite differently in many ways. For dilute
solutions, particles interact via long-range hydrodynamic flows generated by
active particles and the short-range interactions can be safely ignored
\cite{review2}. However, transport properties of dense mixtures are mostly
dominated by the short-range interactions, which are responsible for a
variety of cluster and pattern formation processes reported in the recent
literature \cite{Buttinoni,Fily, Schwarzendahl1, Schwarzendahl2}.

Our simulation of binary active mixtures shows that adding a fraction of
active microswimmers, such as self-propelled JPs, to a suspension of passive
colloidal particles, results in a motility increase of the latter species.
However, adding a small fraction of more active particles to a suspension of
less active microswimmers, results in a non-trivial behavior, whereby the
added species appears to enhance the motility of the host species. Such a
mechanism can be controlled by tuning the parameters ofthe guest species,
e.g., the intensity of light in the case of light-induced JP using laser
beam~\cite{Sano}, near-infrared light~\cite{Lin} or visible light
\cite{Slava1,Slava2}.

Our findings can be potentially useful for various chemical, biological and
medical applications. For example, this technique could be used to increase
the rate of {\it in vitro} fertilization by enhancing the motility of weak
sperm cells. The same strategy can be implemented in a chemical reactor to
govern the reaction rate, whereby one adds inert active particles to stir
otherwise slowly diffusing reactant molecules.

Furthermore, it should be noted that the motility transfer mechanism is
associated with some correlated particle dynamics in the mixture. The most
sophisticated way to quantify motion correlation of weak and strong active
particles is to compute the mixture cooperativity \cite{cooperativity}.
However, we focus here on the role of cooperativity in controlling the
velocity distribution and effusion rates of weak JPs via motility transfer.

The outline of the paper is as follows. In Sec.~\ref{Model} we present a
simple dynamical model for interacting self-propelling JPs in two dimensions,
which we implemented in our numerical simulation code. In Sec.~\ref{Veldis},
we explore the velocity distributions of the two mixture components. We
consider first the case of a mixture of two identical species (single species
case), of noninteracting, Sec.~\ref{VDA}, or interacting particles,
Sec.~\ref{VDB}, and, then, the general case of a binary mixture of two
different species of interacting active JPs. In Sec.~\ref{Effusion}, we
report our data for the effusion rates of the two JP species out of a narrow
opening of the simulation box.
Finally, in Sec.~\ref{Conclusions} we draw a few concluding remarks.

\section{Model} 
\label {Model} 
Let us consider a two-dimensional system consisting of two
types of JPs with different self-propulsion speeds in a thermal bath: $N_w$
with speed $v_w$ and $N_s$ with speed $v_s$.  In the following, we will
refer the subscripts 's' and 'w' to "strong" and "weak" mobility JPs,
respectively. All $N=N_w+N_s$ particles are represented by interacting disks
of radius $r_0$. 
For very short distances they interact with each other via
a truncated Lennard-Jones potential,
\begin{eqnarray}\label{1}
V_{ij} &=& 4\epsilon\left[\left(\frac{\sigma}{r_{ij}}\right)^{12} -\left(\frac{\sigma}{r_{ij}}\right)^{6} \right], \;\; {\rm if}\;\;  r_{ij} \leq r_m \nonumber \\
  &=& 0 \;\; {\rm otherwise},
\end{eqnarray}
where $\epsilon$ is the interaction constant, 
$r_m$ locates the potential minimum, and $\sigma = 2r_0$. 
 Thus,
particles interact only through steric repulsion, i.e., no hydrodynamic
interactions will be considered here. To illustrate how the motility of the
two species are interrelated, we computed two quantifiers, the particles
velocity distributions and their effusion rates. However, the former cannot
be computed for massless particles (that is in the absence of inertia).
Therefore, we assumed damped particle dynamics, although in most practical
situations inertia plays no significant role, due to the comparatively very
fast viscous relaxation of the suspension medium \cite{Gardinar}. One can
recover the standard massless, or overdamped, limit by taking very large
values of the damping constant $\gamma$. This holds on all physical
circumstances when the viscous relaxation time, $1/\gamma$, is much shorter
than any other relevant time scale of the system dynamics
\cite{Our-Intertia,Lowen-inertia1, Lowen-inertia2}.

The dynamics of the particles in the $xy$-plane can be described by the
following set of Langevin equations,
\hspace{-0.5cm}
\begin{eqnarray}
 m \ddot{x_i} &=& -\gamma [\dot{x_i} + \sum_{j} F_{ij}^x +  v_0 \cos{\theta_i} + \sqrt{ D_0} \,\xi_{i}^x (t)],\label{2}\\
m \ddot{y_i} &=& -\gamma [\dot{y_i} + \sum_{j} F_{ij}^y +   v_0 \sin{\theta_i} + \sqrt{ D_0}\,\xi_i^{y} (t)],\label{3}\\
\dot{\theta_i} &=&  \xi_i^{\theta}.\label{4}
\end{eqnarray}

The $i$-th particle with instantaneous position $(x_i, y_i)$ diffuses under the
combined action of self-propulsion and equilibrium thermal fluctuations.
Here, $(\xi_i^x, \xi_i^y)$ are the components of the thermal fluctuations
responsible for the particle translational diffusion; they are modeled by  Gaussian white
noises with $\langle \xi_{i}^{\alpha}(t)\rangle=0 $ and $\langle
\xi_i^{\alpha}(t)\xi_i^{\beta}(0)\rangle=2 \delta_{ij}\delta_{\alpha \beta}
\delta (t)$, where $\alpha, \beta = x, y$. The constant $D_0
= kT/\gamma$ can be computed by measuring the translational diffusion of a
free JP in the absence of self-propulsion. Here
$\gamma$ plays the role of an effective damping constant
incorporating all environmental interactions not explicitly accounted for in Eqs.~(\ref{2})-(\ref{3}),
like fluid viscosity, hydrodynamic drag, surface effects, etc. The
second term in the right hand side of the same equations represents the
repulsive forces derived from the Lennard-Jones pair potential of Eq.~(\ref{1}).

The propulsion velocities with modulus $v_w$ and $v_s$ are oriented at an
angle $\theta_i$ with respect to the laboratory $x$-axis.  Due to the
particles rotational diffusion, the angles $\theta_i$ change randomly
according to the Wiener process of Eq.(\ref{4}), where $\langle
\xi_i^{\theta}(t)\rangle=0$ and $\langle
\xi_i^{\theta}(t)\xi_i^{\theta}(0)\rangle=2D_\theta\delta (t)$. For a passive
particle, the rotational diffusion constant, $D_{\theta}$, is typically
related to the viscosity, $\eta_v$,  and temperature, $T$, of the suspension
medium and to the geometry of the particle itself \cite{Berg}. For spherical
colloidal particles with radius $r_0$, the rotational diffusion constant can
be expressed as $D_{\theta} = kT/8\pi\eta_v r_0^3$. However, for an active
JP rotational diffusion can also depend on the mechanisms fueling its
self-propulsion.  For this reason,  $D_0$, $v_0$, and $D_\theta$ are treated
here as independent model parameters \cite{cataly2,cataly3,Teeffelen}.  Moreover, we assumed for simplicity
that the noise parameters $D_0$ and $D_\theta$ are the same for both JP species.

From the correlation function, $\langle \cos \theta_i (t) \cos {\theta}_i (0)
\rangle = \langle \sin \theta_i (t)\sin \theta_i (0)\rangle =(1/2)\exp[{-D_{\theta}|t|}]$, it is apparent that
$D_\theta$ coincides with the rotational relaxation rate of the
self-propulsion velocity ${\vec v}_0(t)$. Moreover, we remind that, in the
limit of large $\gamma$, a non interacting JP of Eqs. (\ref{2})-(\ref{4})
diffuses normally with translational constant, $D$, consisting of two
distinct terms \cite{our2}, a thermal and a self-propulsion one, namely $D=D_0+v_0^2/2D_\theta$.

\begin{figure*} [h] \centering
\includegraphics[height=0.3\textwidth,width=0.325\textwidth]{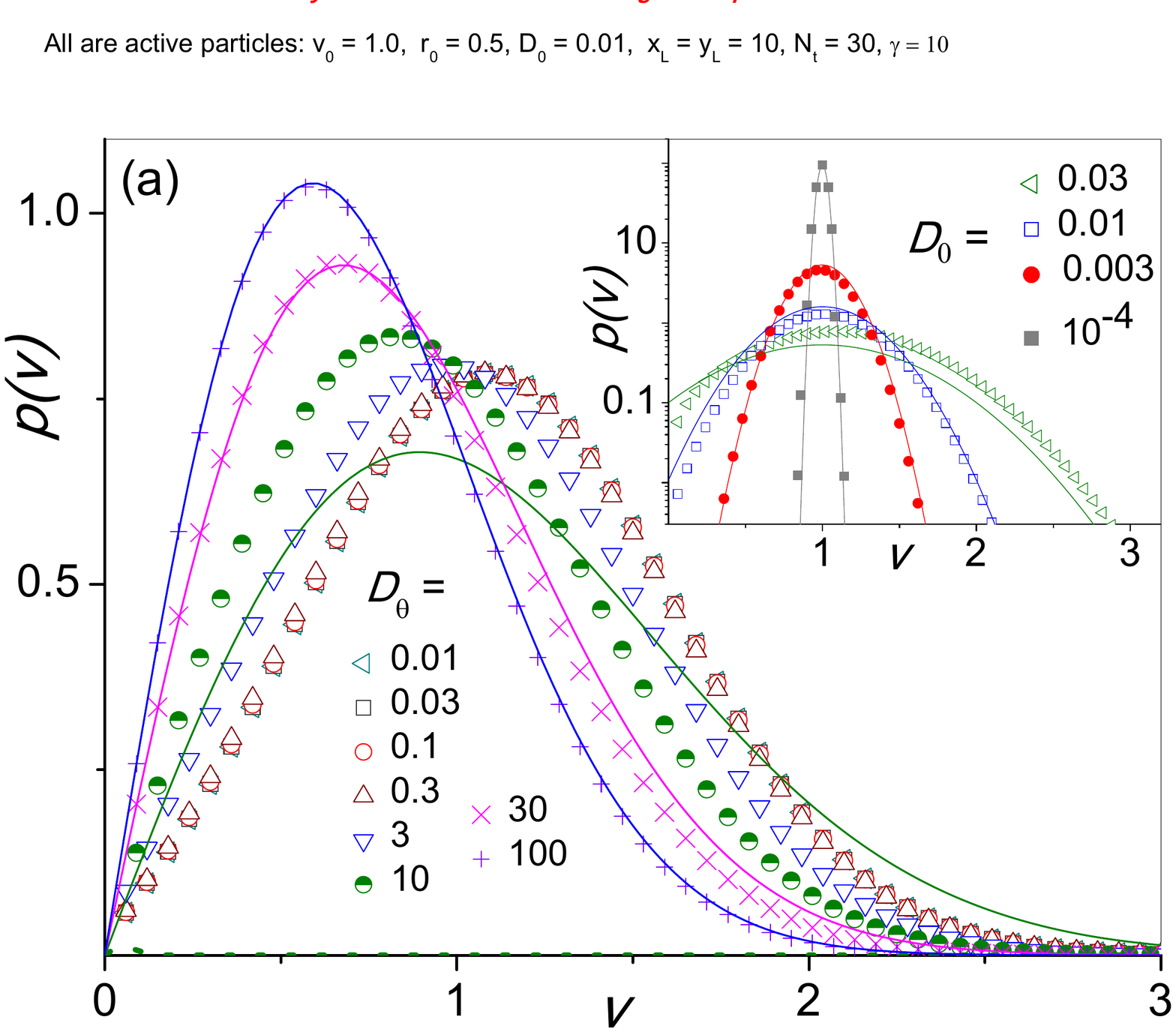}
\includegraphics[height=0.3\textwidth,width=0.325\textwidth]{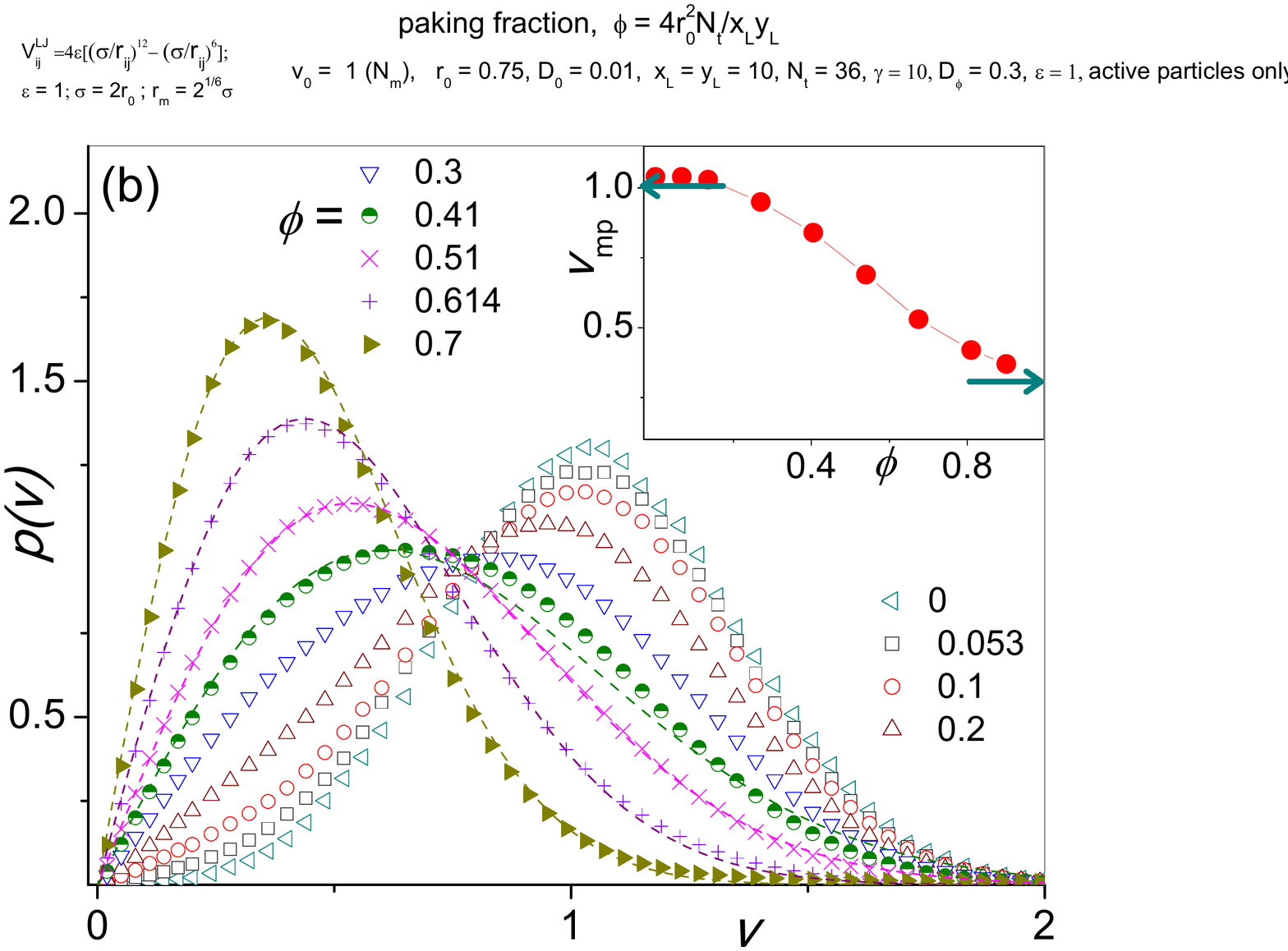}
\includegraphics[height=0.3\textwidth,width=0.325\textwidth]{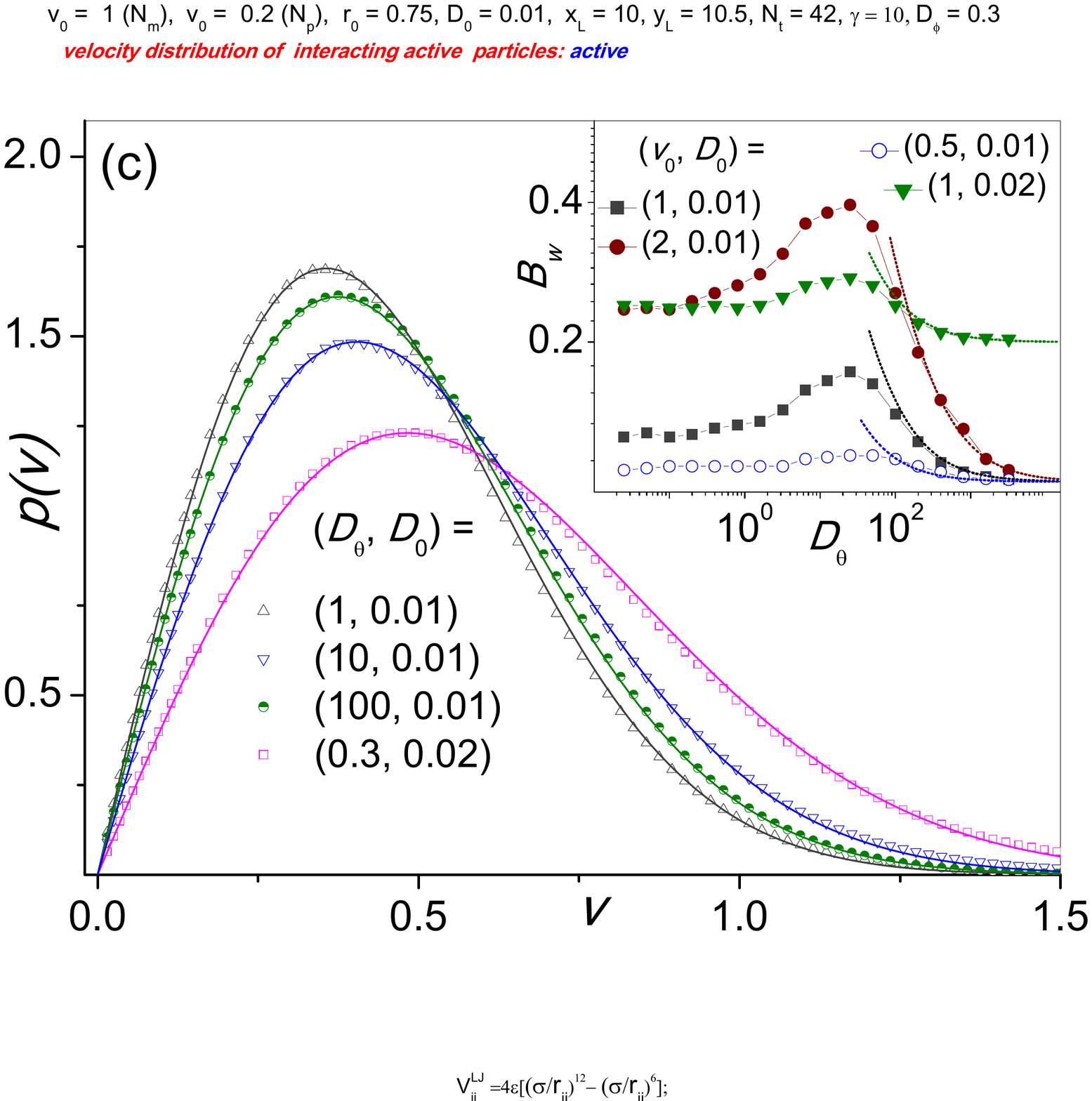}
\caption {(Color online) (a) Velocity distribution of non-interacting active
JPs for different $D_\theta$ (see legends).  The inset illustrates the effect
of thermal noise for $D_{\theta} = 0.3$. Symbols denote numerical simulation
data; solid lines are the analytical estimates of Eq. (\ref{6}), main panel,
and Eq. (\ref{7}), inset.   The parameters used are (unless
reported otherwise in the legends):  $D_0 = 0.03,\; v_0 = 1, \; \gamma =10, \; m =1 $.
(b)Velocity distribution of interacting self-propelled particles
for different the packing fraction, $\phi$.
(c)Velocity distribution of interacting self-propelled particles for
different $D_{\theta}$. In the main panels (b) and (c),
solid lines represent the least-square fitted 2D Gaussian distributions of Eq.(\ref{8}).
 The parameters used are (unless reported otherwise in the legends)):
 $v_0 = 1, \; \tau_{\theta} = 3.33, \tau_{\gamma} = 0.1, r_0 = 0.75, \; D_0 = 0.01,\; \epsilon = 1,\; \phi = 0.7$.
 Insets: (b) most probable velocity, $v_{mp}$, versus packing fraction.
 Asymptotes, at $\phi \rightarrow 0$, $v_{\rm mp} \rightarrow v_0$ and for
 $\phi \rightarrow  1$, $v_{\rm mp} \rightarrow \sqrt{B}$ are depicted  by
 horizontal arrows. (c) Variance of the distribution Eq.(\ref{8}) as a function of
$D_{\theta}$ for different values of  $v_0$ and  $D_0$. Dotted lines are analytic
curves corresponding to the effective temperature of Eq.\~(\ref{5}) (see text).   \label {F2}}
\end{figure*}

We numerically integrated Eqs. (\ref{2})-(\ref{4})  using a standard Milstein algorithm to obtain
the velocity distributions and effusion rates of the both mixture species.
The numerical integration was performed using a very short time
step, $10^{-6}$-$10^{-7}$, to ensure numerical stability.  Computing  the
velocity distributions requires no confinement scheme. However, to keep the mixtures
densities constant, we set up a simulation box of dimension $x_L \times y_L$
with periodic boundary conditions.
Instead, to simulate the effusion rates we assumed that the particles centers are
confined inside the simulation box.  The particles can then exit the box only through a very small
opening of width $\Delta+2r_0$, to model a pore of accessible width $\Delta$ (see Fig. \ref{F1}). The opening
can be centered anywhere along the box wall.  Simulating a confined JP
requires defining its collisional dynamics at the boundaries. For the
translational velocity $\dot {\vec {{r}}}$, we imposed elastic reflection, whereas the
rotational coordinate, $\theta$, was assumed not to change upon collision
(sliding boundary conditions \cite{our1}).  As a consequence, an active JP
tends to slide along the walls until rotational fluctuations, $\xi_{\theta}$,
redirects the particle inside the box.  We computed the effusion
rate, defined as the number of particles exiting the box through the pore per unit of time,  for
different particles swimming properties and confinement geometries. At
$t=0$, the particles were  uniformly distributed  in the box with random
orientation. To keep the number density of both species constant, a particle of the same species was
re-injected with random position and orientation inside the  box,
whenever one had escaped through the pore.  The running time was
set to $10^4 \times \tau_{\theta}$ or $10^4$, whichever was greater, so as to
neglect transient effects due to the initial conditions. The data
points reported in the figures shown here have been obtained by ensemble
averaging over a minimum of 1000 trajectories.
For the simulation parameters values adopted here, the time and length scales are seconds and micrometers, respectively.
The mass of a slilica bead of radius 0.75~$\mu$m is taken as a unit of mass.
Taking the density~\cite{silica} of SiO$_{2}$ $\approx 2$ g/cm$^{-3}$,
the unit of mass would be about $4 \times 10^{-12}$~g.
In rescaled units, parameters used in our simulations are consistent with the corresponding values reported in the experimental literature.

\section{Velocity distribution} 
\label{Veldis}
It is well known that the velocity of overdamped Brownian particles
is an ill-defined quantity. Indeed,
massless particles undergo a displacement
only during the action of external forces~\cite{Gardinar}, here
thermal fluctuations, collisions against other particles or the box walls,
and the effective self-propulsion forces~\cite{JPforce}. Therefore, to extract
a velocity distribution, one needs to simulate inertial effects.
%%%%%

%%%
By numerically integrating the coupled Eqs.~(\ref{2})-(\ref{4}), we systematically
analyzed velocity distributions in systems of non-interacting and interacting
active JPs, as well as in binary mixtures of two species of JPs with different
self-propulsion speeds.

\subsection{Velocity distribution of non-interacting active particles}
\label{VDA} Let us begin with the case of a single species of non-interacting
particles self-propelling in a thermal bath of temperature $T$ with speed
$v_0$. Velocity distribution at different values of the rotational diffusion
constant, $D_{\theta}$, are shown in Fig. \ref{F2}. It is apparent that
inertial effects become important as the viscous relaxation time constant,
$\tau_\gamma=1/\gamma$, grows comparable or greater than the rotational
relaxation time $\tau_\theta = 1/D_{\theta}$. When $\tau_\gamma \gg
\tau_\theta $ (or $\gamma \ll D_\theta$), the velocity  distributions are
mostly determined by thermal fluctuations.  In the opposite regime,
$\tau_\gamma \ll \tau_\theta $ (or $\gamma \gg D_\theta$), self-propulsion
effects seem to prevail. 
 Hence,  the transition from self-propulsion to
inertia-dominated regime, clearly emerging from the velocity distributions of Fig. \ref{F2} 

Recall that, as anticipated above, for asymptotically large observation
times, a free JP behaves like a
persistent Brownian particle with effective temperature\cite{temperature1,temperature2}
\begin{eqnarray}
 T_{\rm eff}=\frac{\gamma}{k}\left(D_0+\frac{v_0^2}{2D_\theta}\right), \label{5}
\end{eqnarray}
and persistence length $l_\theta = v_0 \tau_\theta$.  For suitably large values of
$D_\theta$, the self-propulsion length is shorter than the free thermal length,
$\sqrt{mkT}/\gamma$, that is $T_{\rm eff} \simeq T$. As a consequence, one expects that
the particles velocities must be distributed according to the
two-dimensional Maxwellian function,
\begin{eqnarray}
p(v)=\left(\frac{mv}{kT}\right)  \exp\left(-\frac{mv^2}{2kT}\right). \label{6}
\end{eqnarray}
This assertion is corroborated by the numerical results of  Fig. \ref{F2}(a,b).

A different type of velocity distribution emerges when the self-propulsion
length of the active particle is set much larger than its thermal length. The
ensuing velocity distribution is governed by the self-propulsion dynamics,
its maximum being centered at around $v_0$. Such a distribution results from the
combination of the Maxwellian distribution of Eq. (\ref{6}), and a Gaussian
distribution with mean $v_0$ and variance $kT=\gamma D_0$, both due to
thermal fluctuations.
When lowering the temperature, $T$, the
contribution of the Maxwellian part is quickly suppressed, which results in the 2D Gaussian distribution
\begin{eqnarray}
p(v)=\left(\frac{1}{\sqrt{2\pi \gamma D_0}}\right)  \exp\left[-\frac{(v-v_0)^2}{2\gamma D_0}\right]. \label{7}
\end{eqnarray}
In the zero temperature limit, that is, when translational noise is
negligible with respect to rotational noise, this distribution tends to a
$\delta$-function centered at $v_0$, whereas the corresponding velocity
distributions in one direction become, $p(v_{x,y}) = 1/\pi \sqrt{1 -
(v_{x,y}/v_0)^2}$.  These properties are confirmed by the simulation
results displayed in Fig. \ref{F2}(a) and supplementary FigSM1.

\subsection{Velocity distribution in a system of interacting active particles}
\label{VDB}
Velocity distributions for different values of the packing fraction, $\phi=4r_0^2N_t/(x_L+2r_0) (y_L+2r_0)$,
are displayed in Fig. \ref{F2}(b).  These distributions are centered at
$v_0$ for weakly interacting particles and their center shifts towards lower
values with increasing $\phi$. As apparent here, in dense systems, say with
$\phi > 0.5$, interacting active JPs obey the Maxwellian
velocity distribution,
\begin{eqnarray}
p(v)=\frac{v}{B} \exp\left[-\frac{v^2}{2B}\right], \label{8}
\end{eqnarray}
where the fitting parameter, $B$, depends on the bath temperature $T$, the particles
rotational diffusion $D_{\theta}$ and self-propulsion speed $v_0$, and the system packing fraction $\phi$.
For $v_0 \rightarrow 0 $, the distribution is insensitive to the pair interaction,
so that $B = \gamma D_0$, like in gas kinetic theory. Therefore, the
interaction dependence of the velocity distribution is a non-equilibrium
effect of self-propulsion. To examine the impact of
self-propulsion on the velocity distribution in a dense system, in Fig. \ref{F2}(c) we
plotted $p(v)$ (main panel) and distribution width $B$ (inset) as a
function of $D_{\theta}$ for different values of the speed $v_0$. One notices
immediately that:
\newline (i) For very slow rotational relaxation, the width of the
distribution is almost independent of $D_{\theta}$. In this regime, the
self-propulsion length $l_{\theta}$ is much larger than the average
effective inter-particle distance $l_s$, so that the particles free path
cannot exceed $l_s$. The fitting parameter seems to obey the empirical law,
$B = \gamma \left(D_0 + \alpha  v^2 \right)$, with $\alpha$ a function of the
packing  fraction.  This result can be explained by comparing $B$ with
$kT_{\rm eff}$ in Eq. (\ref{5}), which we rewrite here as $kT_{\rm
eff}=\gamma(D_0+v_0^2\tau_\theta/2)$. Upon increasing $\phi$, both the mean-free ballistic time
$\tau_s=l_s/v_0$ and the mean-free diffusion time $\tau_D=l_s^2/2D_0$ grow
larger than the persistence
time $\tau_\theta$. As a
consequence, $\tau_\theta$ in the above expression for $kT_{\rm eff}$ should now be
replaced by $\overline \tau={\rm min}\{\tau_s,\tau_D\}$. When the
active suspension is so dense that $D_0>v_0l_s/2$, then $\overline \tau=\tau_D$,
so that the fitting parameter $B$ depends quadratically on $v_0$ with $\alpha$ a
function of $\phi$.

On a closer look, one notices that $\alpha$  also weakly depends
on $v_0$. This is because self-propulsion makes the colliding particles to occasionally overlap,
thus slightly lowering the effective $\phi$ value. The pair penetration length and, hence,
the effective particles size, can be estimated by equating the self-propulsion
force to the inter-particle repulsion.
\newline (ii) In the opposite limit,
$l_\theta <l_s$, the active particles manage to change their direction before
colliding with other particles, so that their
inter-collisional dynamics is dominated by the self-propulsion dynamics.
They behave as if they were floating in a thermal bath with the effective
temperature of Eq. (\ref{5}). The ensuing estimates of the distribution
fitting parameter,  $B = \gamma (D_0 +v^2\tau_\theta/2)$, drawn in the inset of Fig. \ref{F2}(c)
fairly agree with the numerical data.

Figure \ref{F2}(b,c) [and the supplementary figure FigSM2] also suggests that, under the condition that $v_0^2 \gg B$,
the most probable value of $v$, $v_{\rm mp}$, approaches $v_0$. Based on our
argument of (i) for dense active suspensions, this requires $v_0^2 \gg \gamma
D_0/(1-\alpha)$, with $\alpha=l_s^2\gamma/4D_0$. Of course,
this estimate holds only for not too large $\gamma$ values,
so that $\alpha < 1$, i.e., for $l_s^2\gamma/4 < D_0$.
\newline (iii) In the intermediate regime, the curves $B$ versus $D_\theta$ exhibit a
maximum. Starting with $l_s \ll l_\theta$, as one increases the rotational
diffusion constant, self-propulsion enters gradually into play by {\it
enhancing} $B$. On the other hand, self-propulsion effects disappear in the
diffusive regime, $l_\theta \ll l_s$, where $B$ decreases with  increasing
$D_{\theta}$. Not surprisingly, $B$ appears to reach its maximum in the
intermediate regime for $l_\theta$ of the order of the mean inter-particle
distance $l_s$.

Finally, it should be noted that the fitting values of $B$ have been
extracted by least-square fitting. The fidelity of such fittings has been
assessed by computing the mean square weighted deviation~\cite{least-square}
$\chi^2_v$. It always returned values close to 1, except for large $v_0$.
This deviation is noticeable for $v_0 = 4$, where the rotational diffusion is
rather low (shown in supplementary FigSM2).

%\begin{figure*}
\begin{figure*}[h]
\centering
\includegraphics[height=0.3\textwidth,width=0.325\textwidth]{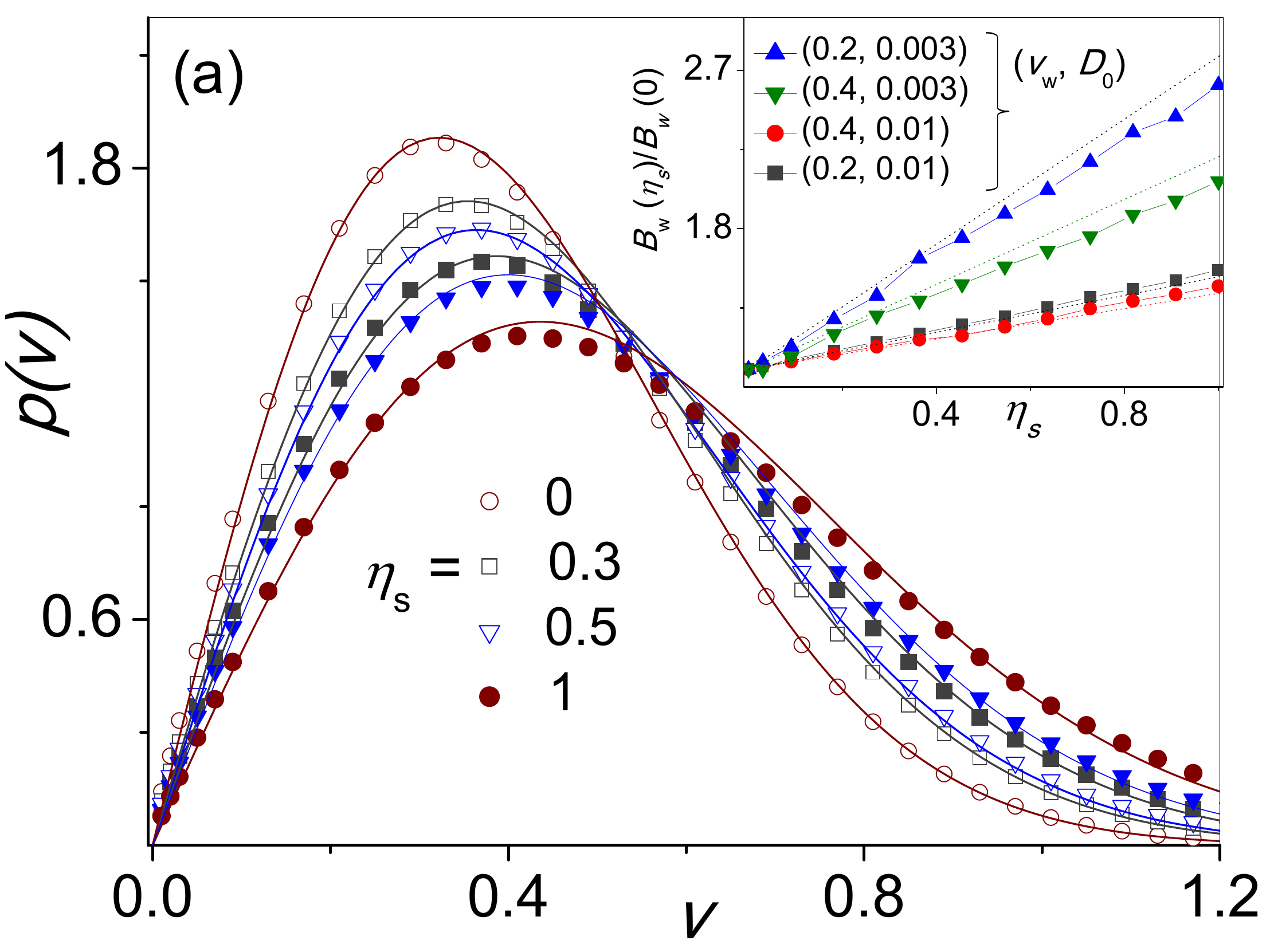}
\includegraphics[height=0.3\textwidth,width=0.325\textwidth]{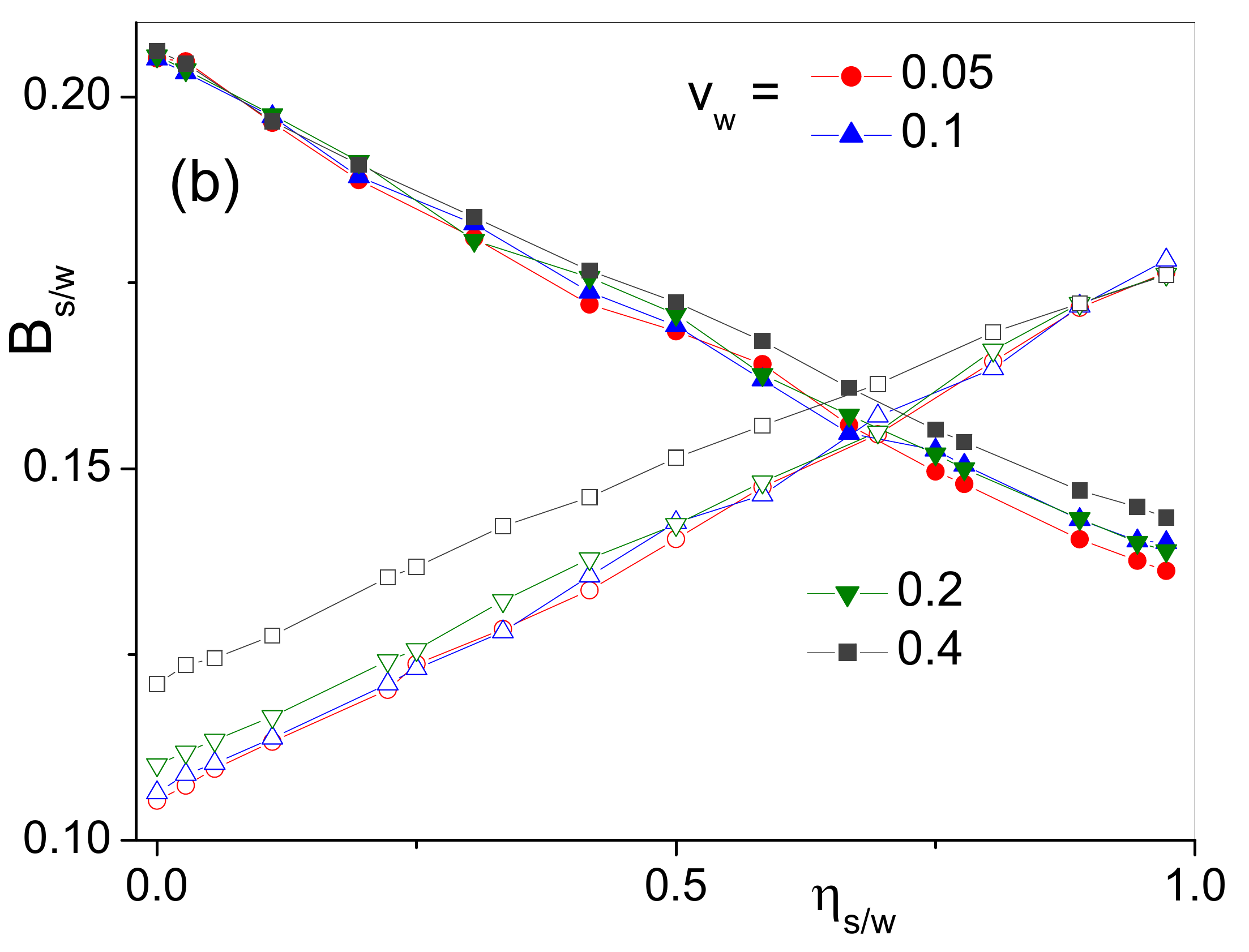}
\includegraphics[height=0.3\textwidth,width=0.325\textwidth]{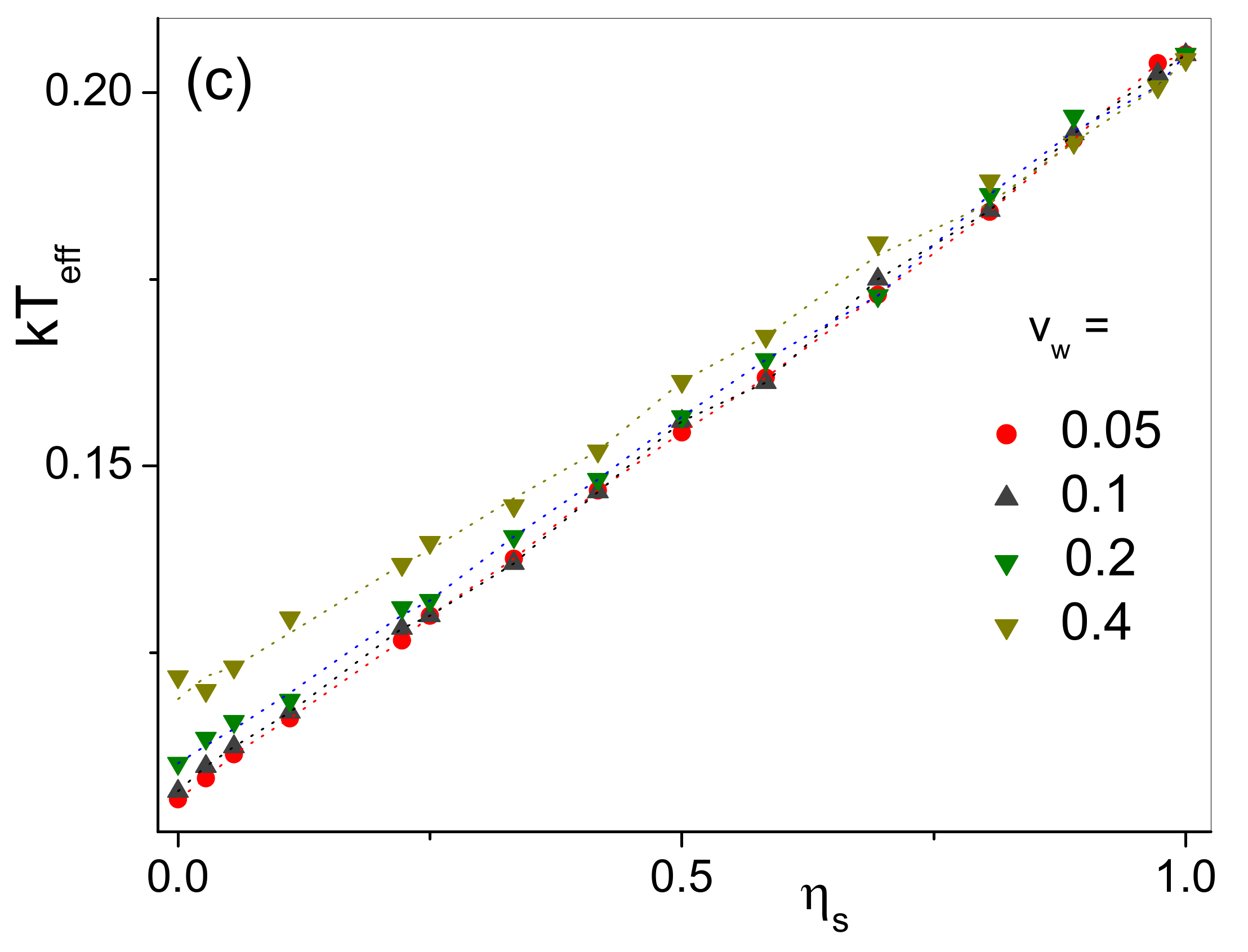}
\caption {(Color online) (a) Comparison of velocity  distribution $p(v)$ of
weak (empty symbols) and strong (filled symbols) active particles with
varying composition $\eta_s$ of the binary mixture. Note that solid (hollow)
circles represent velocity distribution of strong (weak) active particles for
the interacting single species case. Solid lines correspond to Eq.~(\ref{8})
where $B$ is obtained from least square fittings. Inset plots depict the
variation of distribution width of weak JPs $B_w$ as a function of the
fraction of strong active particles, $\eta_s = N_s/N_t$. 
Thus, 
$\eta_s = 0$ means that all particles are weak and 
$\eta_s = 1$ that all particles are strong. 
Dotted lines
represent Eq.(\ref{9}) with the relevant best-fit parameters $\alpha_s$ and $\alpha_w $. (b)
Distribution widths of weak, $B_w$ (empty symbols),
and strong active JPs, $B_s$ (filled symbols), respectively
versus $\eta_s$ and $\eta_w$ for different $v_w$.   (c) $kT_{\rm eff}$  versus $\eta_s$
for different $v_w$: numerical data (symbols) are compared with the analytical
estimates of Eq. (\ref{Tm2}) (dotted lines). The remaining model parameters are (unless
reported otherwise in the legends): $v_s = 1, \; D_{\theta} = 0.3,  \tau_{\gamma} =
0.1, r_0 = 0.75,\; D_0 = 0.01, \; \epsilon = 1,\; \phi = 0.61$.
 \label{F3}}
\end{figure*}
%\end{figure*}

\subsection{Velocity distribution in a binary mixture of active particles}\label{VDC}

Let us consider now a mixture of active particles of two types. Let us denote
the $N_w$ particles with fixed self-propulsion speed $v_w$, as {\it weakly}
active, and the remaining $N_s$ particles with tunable self-propulsion
velocity $v_s$,  as {\it strongly} active. A comparison of velocity
distributions of the weak (hollow symbols) and  strong (solid symbols) JPs in
a binary mixture are shown in Fig. \ref{F3}(a) for different fractions
$\eta_s=N_s/N$ of the strong active particles. Plots here correspond to
situations where the system packing fraction $\phi$ is quite large and the
velocity distributions $p(v)$ are of the Maxwellian type, Eq. (\ref{8}).
 As to be expected, the plots in Fig. \ref{F3}(a) show that weak active
particles distributions grow wider, and their maxima shift to higher velocities, with increasing $\eta_s$. 
On the other hand,  the distributions of the stronger component shrink and their maxima shift toward lower velocity values in comparison with the single component system. 
This result suggests an effective {\it motility transfer} from more active to
less active particles. To better characterize the underlying mechanism, we
estimated the distribution half-widths $B$ for different mixture
compositions. The ratio $B_w(\eta_s)/B_w(0)$ in the inset of Fig. \ref{F3}(a)
grows linearly with $\eta_s$, its slope depending on the thermal energy,
$kT=\gamma D_0$, and self-propulsion speed of both JP species.  This behavior
can be explained as follows. Since the system is dense and $l_s \ll
l_{\theta}$, self-propulsion only contributes to the effective thermal motion
of the system, see item (i) of Sec. \ref{VDB}. Adding up the average kinetic
energy contribution from both species and equating the result to the
corresponding prediction based on Eq.(\ref{8}), one can arrive at
\begin{eqnarray}
\frac{B_{i}(\eta_{j})}{B_{i}(0)} = 1+ \left( \frac{ \alpha_j v_j^2-\alpha_i v_i^2 }{ \gamma D_0 + \alpha_i v_i^2} \right)  \eta_{j}.  \label{9}
\end{eqnarray}
Where, $\{i,j\} = \{ s,w \}$ with $i\neq j$. The above estimate
rests on the assumption that the self-propulsion contributions to the kinetic
energy in this regime are directly proportional to $v_{i}^2$ with a
proportionality constant $\alpha_i$. For $l_s \ll l_{\theta}$, both
$\alpha_i$ are insensitive to the rotational diffusion constant $D_\theta$,
and weakly depend on $v_i$. On the contrary, for $l_s \gg l_{\theta}$,
$\alpha_w = \alpha_s = \gamma/2D_{\theta}$.

 To better interpret the mechanism of host-guest mobility transfer, in
Fig. \ref{F3}(b) we compare the widths $B_{s}$ and $B_w$ of the relevant
velocity distributions. We simplify our analysis by focusing on the parameter
regimes where both  mixture components exhibit a Maxwellian velocity distribution. 
Figure \ref{F3}(b) shows that $B_w$ linearly grows as the fraction, $\eta_s$, of strong active particles increases. 
By contrast, $B_s$
decreases with increasing $\eta_w$. 
Equation (\ref{9}) is useful to explain the linear dependence of both $B_s$ and $B_w$ on $\eta_s$. 
It is apparent from both numerical simulations and Eq.~(\ref{9}) that in a binary mixture the velocity distribution of the weak host depends not as much on its own self-propulsion parameters as  on the presence of the strong guest. 
Under the Maxwellian conditions assumed here, $v_w^2 \ll v_s^2 $ and $\gamma D_0 \gg \tau_{\theta}v_w^2 $, one can easily relate the effective temperature of the binary mixture to the distribution widths $B_{s,w}$ as follows, 
\begin{eqnarray}
 kT_{\rm eff} = (1-\eta_s) B_w(\eta_s) + \eta B_s(\eta_s) \label{Tm2}.
\end{eqnarray}
This estimate for $T_{\rm eff}$ is in good agreement with the numerical
results shown in Fig.~\ref{F3}(c). 
In view of the linear $\eta$ dependence of $B$, one would then expect $T_{\rm eff}$ to be a nonlinear function of $\eta$. However, by inspecting Eqs.(\ref{9})-(\ref{Tm2}) one easily concludes that, for the simulation parameters adopted in 
Fig.~\ref{F3}(c), nonlinear corrections are negligible. 
As a result, the effective temperature of the binary mixture grows (almost) linearly with the mole fraction of the guest particles. 
Moreover, Eq.~(\ref{9}) also hints at how the self-propulsion properties of the host and guest particles impact $T_{\rm eff}$.

%Even in case of low packing fraction, velocity distribution of weak active particles (not shown) considerably improved by the guest particles.  Simulation results show that in this regime analysis based on the framework of Maxwellian  distribution does work. However, effective temperature of the system linearly grows with mole fraction of more active guest particles. In the limit, $l_s > l_{\theta}$ effective temperature can be estimated based on the equation (5).}

\begin{figure} [h]
\centering
\includegraphics[height=0.25\textwidth,width=0.45\textwidth]{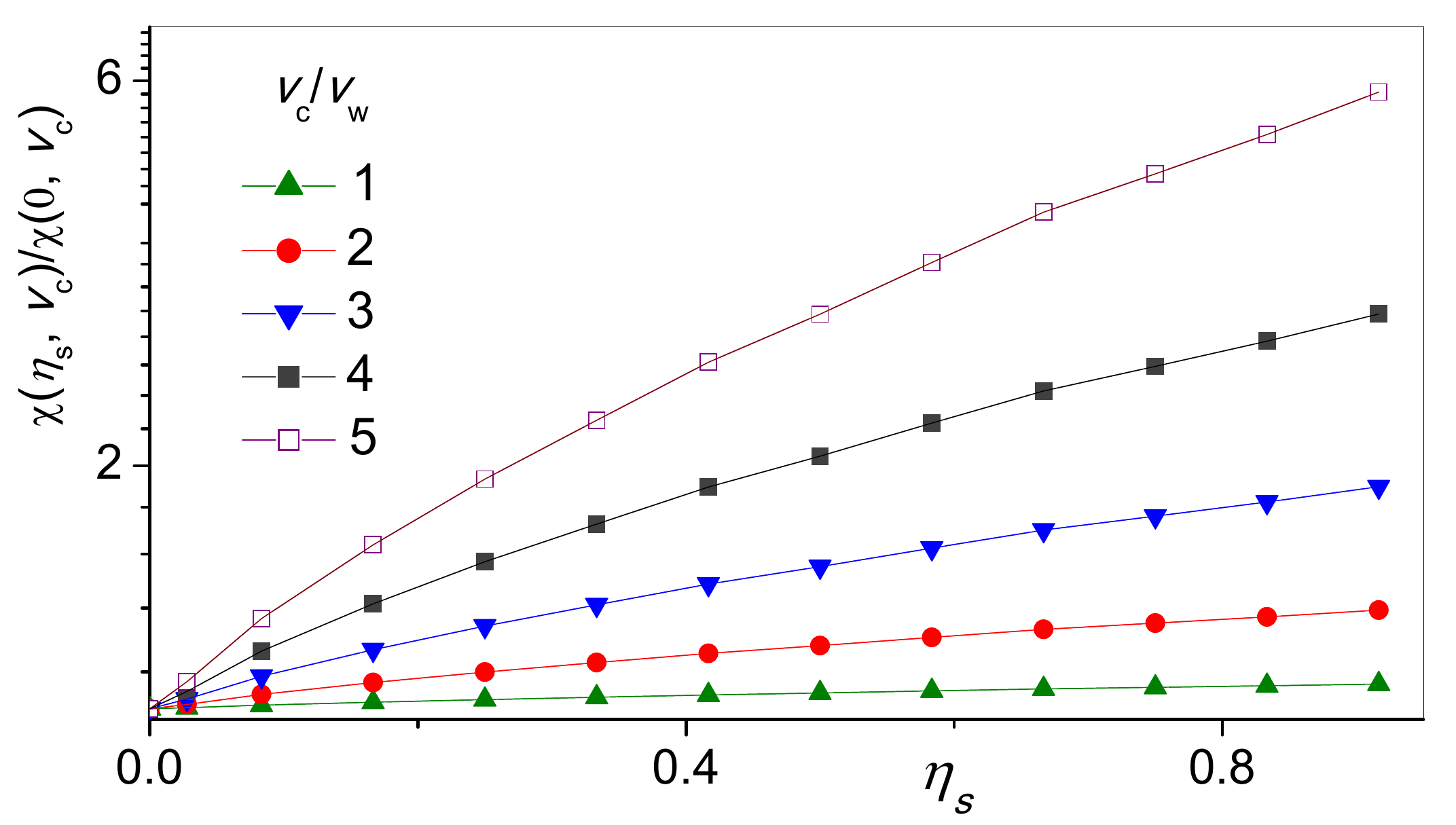}
\caption { (Color online) The ratio, $\chi(\eta_s,v_c)/\chi(\eta_s,v_c)$ versus $\eta_s$ for different $v_c$.
The parameters used are (unless mentioned in the legends): $v_s = 1, \; v_w = 0.2,\; D_{\theta} = 0.3,  \tau_{\gamma} = 0.1, r_0 = 0.75,\; D_0 = 0.01, \; \epsilon = 1,\; \phi = 0.8$ .
 \label{F4}}
\end{figure}

One often needs to know the fraction of {\it weakly} active particles whose
speed exceeds a specified value,  say $v_{c}$. One can calculate this
quantity, $\chi(\eta_s, v_c)$, directly from the velocity distribution
function of the less active JPs, that is
\begin{eqnarray}
\chi(\eta_s, v_c) =  \int_{v_c}^{\infty} p(v,\eta_s)\;dv.  \label{10}
\end{eqnarray}
Thus, $\chi(\eta_s, v_c)$ is the fraction of {\it weakly} active particles
having  an instantaneous velocity greater than the cut-off velocity $v_c$ in
the binary mixture with $N_w$ weak JPs.
 To clarify the role of $v_c$, we consider the kinetic model of reaction rate theory. 
As the reactant particles collide with each other, only
a certain fraction of such collisions leads to the formation of the desired
product. To this purpose it is necessary that the energy of the reactants at the
moment of the impact exceeds a threshold value, $E_a$, also known as reaction
{\it activation energy}, which corresponds the the cut-off activation speed,
$v_c = \sqrt{2mE_a}$. Therefore, coming back to the problem at hand, it would
be desirable to know how the ratio $\chi(\eta_s, v_c)/\chi(0, v_c)$ changes
by adding a certain amount of strongly active particles.
For the velocity distributions of Eq.~(\ref{8}), such ratio reads
\begin{eqnarray}
\frac{\chi(\eta_s, v_c)}{\chi(0, v_c)} =  \exp\left[ -\frac{v_c^2}{2}\left(\frac{1}{B(\eta_s)} - \frac{1}{B(0)}\right)\right].  \label{10}
\end{eqnarray}
This quantity, namely the ratio of the number of weakly active JPs with speed
larger than $v_c$ to the same number, but in the absence of strongly active
JPs, is plotted in Fig.~\ref{F4} for different values of $v_c$. Our
simulations show that $\chi(\eta_s, v_c)/\chi(0, v_c)$ is a monotonically
growing function of $\eta_s$; its growth rate increases with increasing $v_c$.
These observations support the strategy discussed in Sec.~\ref{Intro} aiming
at enhancing the motility of weakly active, or even passive particles, by
adding to the system a small fraction of strongly active particles as
autonomous stirrers.

\begin{figure} [h] \centering
\includegraphics[height=0.335\textwidth,width=0.415\textwidth]{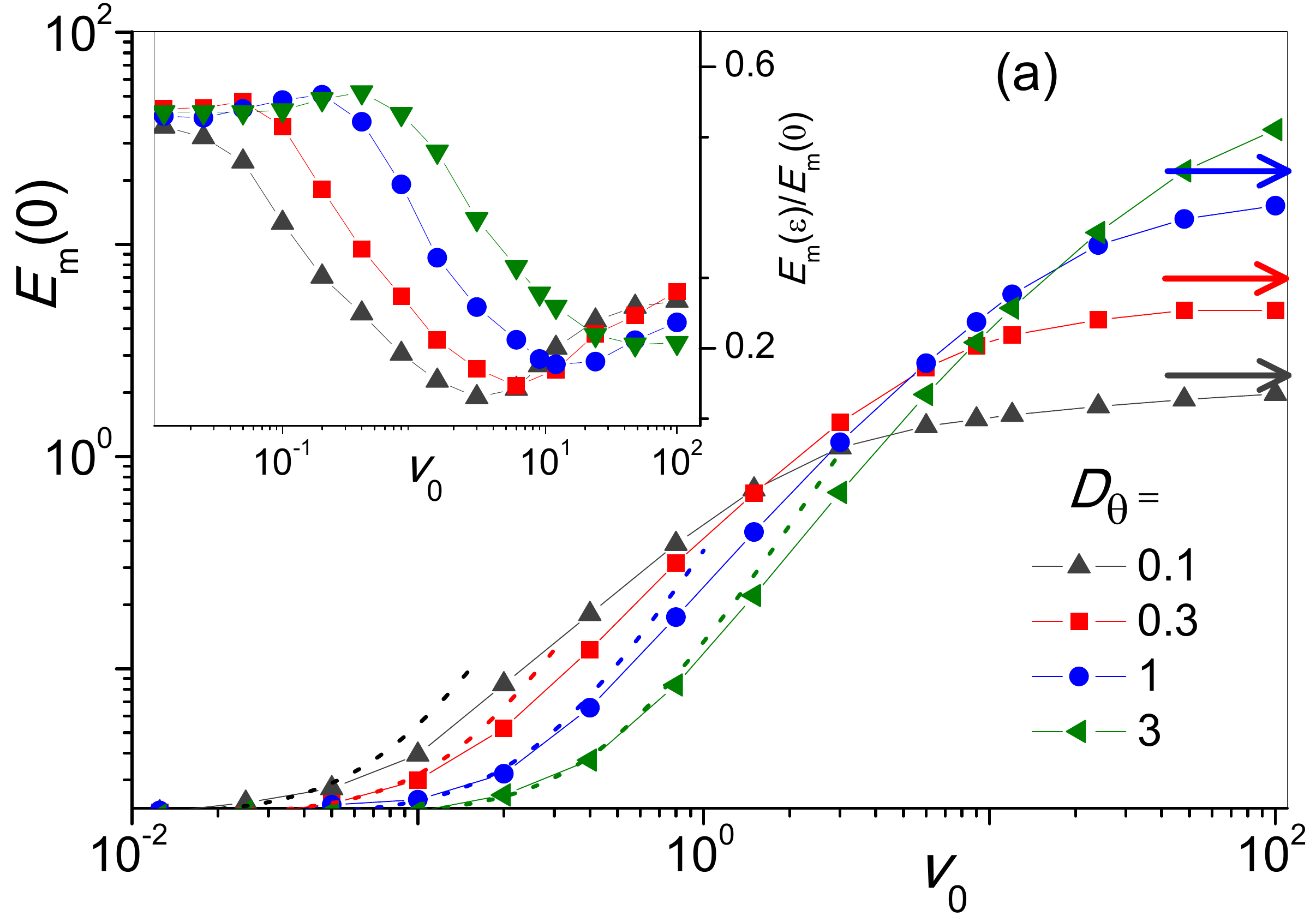}
\includegraphics[height=0.335\textwidth,width=0.415\textwidth]{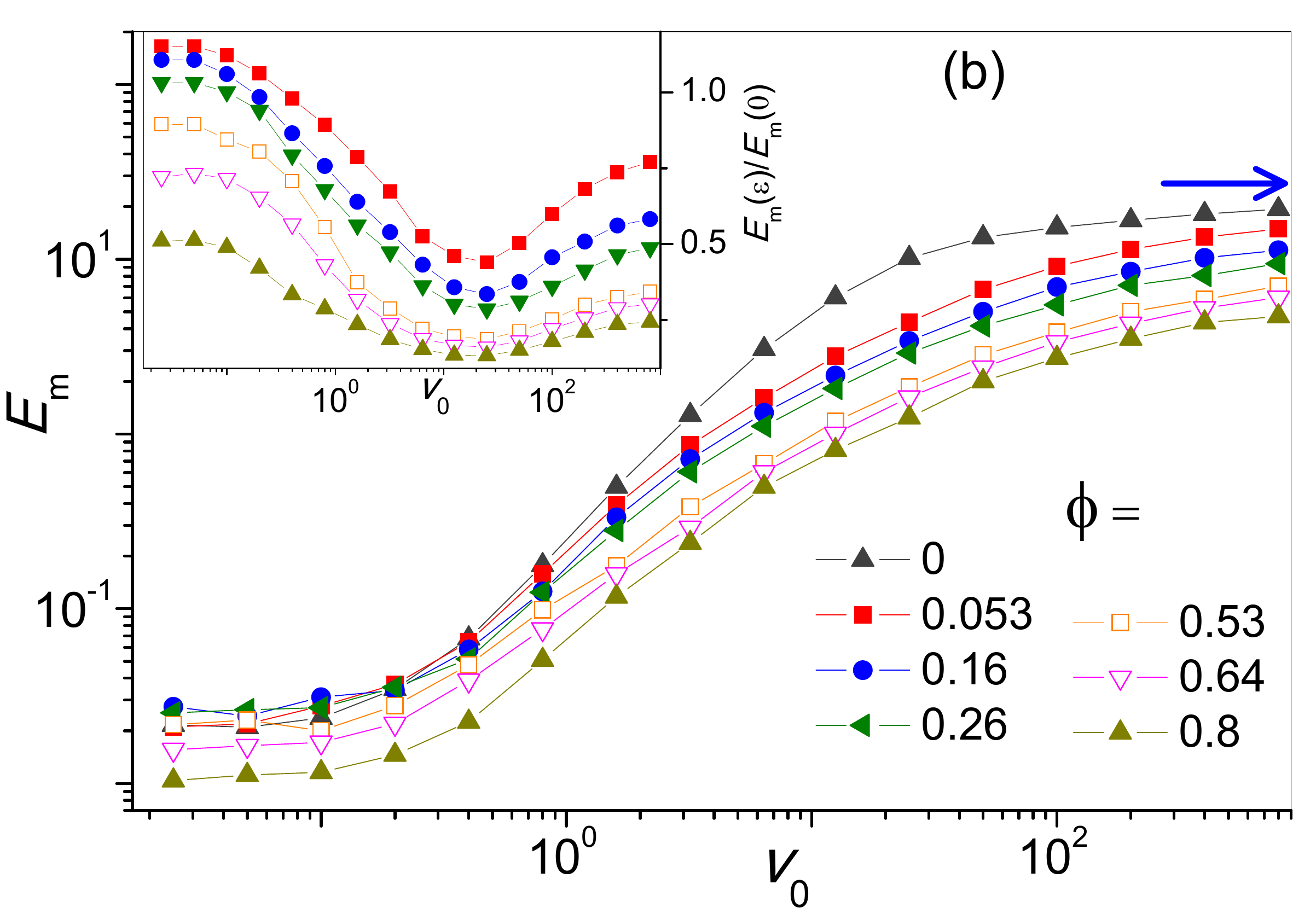}
\caption { (Color online)
(a) Effusion rate $E_m(0)$ of non-interacting JPs with $\epsilon=0$,
as a function of the self-propulsion velocity $v_0$  for different
rotational diffusion coefficient $D_\theta$.
Dotted lines are the predictions based on Eq.~(\ref{12}). Horizontal
arrows indicate the corresponding rate upper bound,
Eq.~(\ref{13}), for large $\tau_\theta=1/D_\theta$.
Inset: the effusion rate ratio $E_m(\epsilon)/E_m(0)$ for $\epsilon = 0.1$
and different $D_\theta$ (see leends). (b) Effusion rate $E_m(\epsilon)$ of
interacing self-propelled particles versus $v_0$ for $\epsilon = 1$ and
different packing fraction $\phi$. Inset:  $E_m(\epsilon)/E_m(0)$ versus
$v_0$ for the same set of parameter as the main panel. Other simulation
parameters for main panels and inset: $x_{L} = y_{L}=10, \; \Delta = 0.5, r_0
= 0.5,\; D_0 = 0.03,\; N_{m} = 80$.
 \label{F5}}
\end{figure}

\begin{figure} [h] \centering
\includegraphics[height=0.335\textwidth,width=0.415\textwidth]{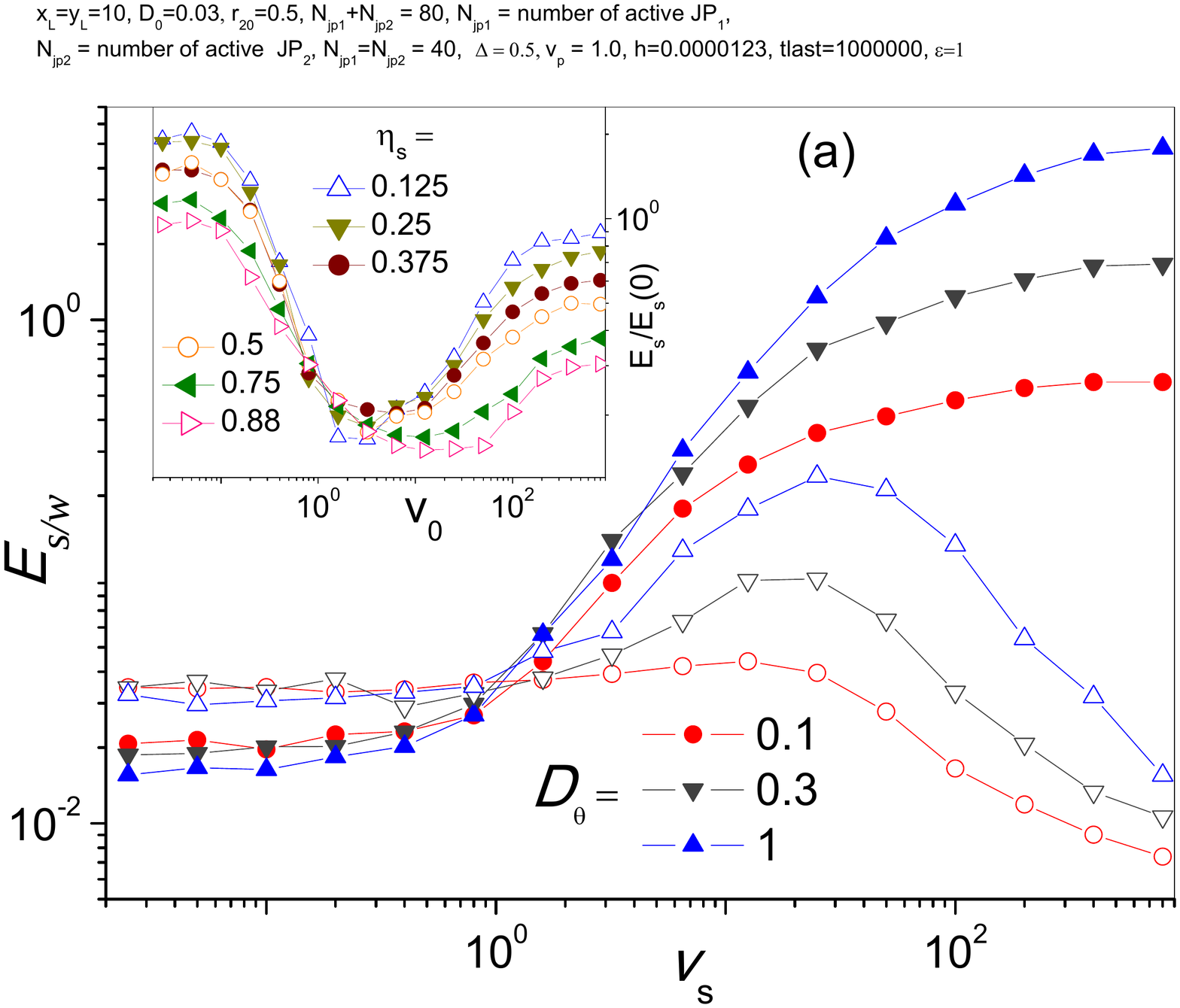}
\includegraphics[height=0.335\textwidth,width=0.415\textwidth]{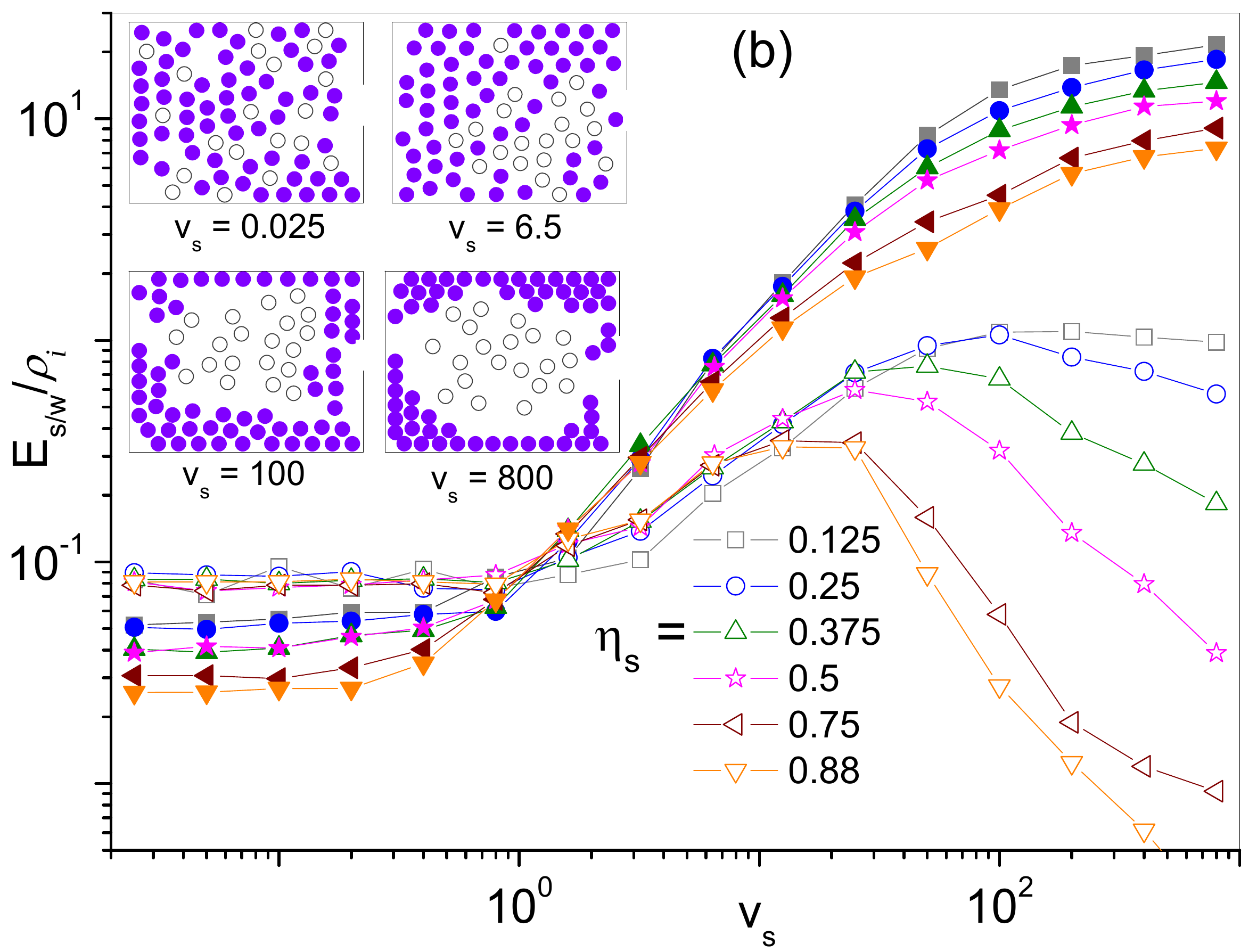}
\caption { (Color online)
(a) Effusion rates $E_s$ (filled dots) and $E_w$ (empty dots) versus
$v_{s}$ for binary mixture with $\eta_s=0.5$ and different values of $D_\theta$ (see legends).
Inset: Effusion ratio of stronger component, $E_m(\epsilon)/E_m(0)$ versus
$v_0$ for different $\eta_s$ and $D_\theta = 1$.
(b)  Effusion rates $E_s$ (filled dots) and $E_w$ (empty dots) versus
$v_{s}$ in a binary mixture for $D_\theta =1$ and different
$\eta_s$ (see legends). Inset: Snapshot of binary
mixture with $\eta_s = 0.75$, $D_\theta = 1$ and different $v_s$  [$0.025$ (top-left),  $6.5$ (top-right),
$100$ (bottom-left),  and $800$ (bottom-right)]. Filled and empty  circles represent strong and weak JPs, respectively. 
Other simulation parameters for main panels and insets:
$v_w =1, \; \epsilon = 1'\; x_{L} = y_{L}=10, \; \Delta = 0.5, r_0 = 0.5,\; D_0 = 0.03,\; N_{m} = 80$ \label{F6}}
\end{figure}

\section{Effusion} 
\label{Effusion} 
In the previous section, we showed how adding a relatively
small fraction of highly motile microswimmers to a suspension of less motile
microswimmers can considerably enhance the overall motility of the mixture.
This effect was demonstrated in the presence of inertia. We consider now the
limiting case of overdamped, or massless, active particles. This limit
corresponds to low Reynolds numbers, a hydrodynamic regime that applies to
most microswimmers investigated in the literature, both biological and
artificial. This raises a problem, because, as mentioned above, the velocity
distribution of massless particles is mathematically ill-defined. To avoid
this difficulty, in our simulations we computed an alternative motility
quantifier for the overdamped limit, namely the effusion rate of the active
JPs through a narrow pore of the simulation box. The corresponding Langevin
equations in the highly damped situation are obtained by ignoring inertia
in Eqs.~(\ref{2})-(\ref{4}),
\begin{eqnarray}
 \dot{x_i} &=& \sum_{j} F_{ij}^x +   v_0 \cos{\theta_i} + \sqrt{ D_0}\,\xi_{i}^x (t),\label{12}\\
\dot{y_i} &=& \sum_{j} F_{ij}^y +   v_0 \sin{\theta_i} + \sqrt{ D_0}\,\xi_i^{y} (t),\label{13}\\
\dot{\theta_i} &=& \sqrt{D_{\theta}} \,\xi_i^{\theta}.\label{14}
\end{eqnarray}
The effusion rate has been studied in depth to characterize classical transport in
constrained geometries  \cite{singer}.
We define the effusion rate of the {\it strong} ($s$)
[{\it weak} ($w$)] JPs, $E_s$ ($E_w$), as the number of $s$ ($w$) particles exiting the simulation box per unit time.
In the case of a single-component system, we denote the effusion rate by $E_m$.

Let us consider the effusion rate $E_{m}(0)$ of a single species of
non-interacting JPs with $\epsilon =0$. In Fig.~\ref{F5}(a) we plotted a few
curves $E_{m}(0)$ versus $v_0$ for different values of $D_{\theta}$. For $v_0
\rightarrow 0$, the effusion is controlled by thermal motion and, as
expected, is insensitive to $v_0$. Effects due to self-propulsion become
appreciable only for values of $v_0$ larger than the particles thermal speed
$\sqrt{2D_0 D_\theta}$. Beyond this critical value, the effusion rate grows
first quadratically with $v_0$ and then saturates toward an asymptotic value.
The rising branches occur for $l_{\theta} \ll x_L, y_L $. Indeed, for very
short rotational relaxation times $\tau_\theta$, when it can safely be
assumed that particles diffuse in a thermal bath with effective constant
$D_{\rm eff}$, the effusion rate through a narrow pore of effective width
$\Delta \ll x_L,y_L$,  reads \cite{singer,jcp1,jcp2}
\begin{eqnarray}
E_{i} =  \pi \rho_i D_{\rm eff}  \left[ \ln\left( \frac{x_L+y_L}{\Delta} \right) \right]^{-1}. \label{12}
\end{eqnarray}

Here, the suffix $i$ refers to either $s$ or $w$, $\rho_i$ denotes the
number density of the mixture component $i$, and $D_{\rm eff}$ is now $D_0+v_0^2/2D_\theta$ --
see Eq.(\ref{5}). This estimate for $E_i(0)$ agrees fairly closely
with the simulation results reported in Fig. \ref{F5}(a). In the opposite
rotational regime, when $l_{\theta} \gg x_L, y_L$, the slow direction changes of the
self-propulsion velocity tends to suppress the particles effusion through the pore. Assuming
that $\tau_\theta $ is much larger than any other system time scale, the
effusion rate can be approximated by \cite{jcp1}
\begin{eqnarray}
E_{i}(0) \approx  { x_L y_L \rho_i D_{\theta}}/{\pi}. \label{13}
\end{eqnarray}
This asymptotic estimate has been marked in Fig. \ref{F5}(a) by horizontal arrows.

The effusion rate of interacting self-propelling particles with $\epsilon >0$
is plotted in the inset of Fig. \ref{F5}(a). This figure shows the
$v_0$-dependence of the effusion rate relative to the corresponding rate in
the absence of interaction,  $E_m (\epsilon)/ E_m (0)$, for several values of
$D_\theta$. The system we simulated here was quite dense ($\phi = 0.66$), so
that the self-propulsion mechanism becomes strongly constrained, being $l_s
\ll l_\theta$. Like in non-interacting systems, the effusion rate is
insensitive to self-propulsion with low $v_0$. More remarkably, with
increasing $v_0$ the relative effusion decreases.

 We attributed this result to the jamming of the interacting particles
caused by self-propulsion in the vicinity of the box walls. Snapshots of the
mixture configurations [see inset of Fig \ref{F6}(b) and supplementary
FigSM3] corroborate this assertion. 
The jamming effect becomes noticeable as soon as the self-propulsion length becomes larger than the confining box. 
Therefore, the appearance of such an effect and minima of
$E_m(\epsilon)/E_m(0)$ versus $v_0$ are inversely related to the rotational diffusion [see inset of Fig. \ref{F5}(a)]. By the same token, one expects that both the decaying and raising branches of the curves $E_m (\epsilon)/ E_m
(0)$ versus $v_0$ are quite insensitive to the packing fraction, $\phi$, in
agreement with the data plotted in the inset of Fig. \ref{F5}(b). In the very
strong self-propulsion regime, both $E_m (\epsilon)$ and $ E_m (0)$ tend to
saturate [see Fig. \ref{F5}(b)]. However, $E_m(\epsilon)$ saturates at larger
$v_0$ values than in the non-interacting case. A plausible explanation is
suggested by a comparison of the mixture snapshots. The particles far away
from the walls are more mobile and contribute more to the effusion rate; they
are not jammed against the walls and ``see'' a larger opening-width to
compartment-size ratio, $\Delta/y_L$. In contrast, particles jammed against
the walls tend to clog the box opening. However, the fraction of the more
mobile particles drops fast with increasing $v_0$, thus leading to plateaus in the effusion rate in the limit $v_0 \rightarrow \infty$.

 Figure \ref{F5}(b) shows that the clogging mechanism works even at low
packing fraction, though its impact on effusion is reduced. 
More remarkably, the excluded volume effect become apparent for $v_0 \rightarrow 0$: the interacting particles become more effusive than the non-interacting ones. 
In a dilute solution, this effect persists until the self-propulsion length grows larger than the average inter-particle spacing. This explains why, in the presence of strong self-propulsion, the computed effusion ratios still grow with $v_0$, though quite slowly.

Figure \ref{F6}(a) illustrates the dependence of the effusion rates of
the two active mixture components on their self-propulsion parameters, $v_0$
and $D_\theta$. The mixture is of 1:1 molar ratio of  strongly ($s$) and
weakly ($w$) active particles. We kept the self-propulsion speed $v_w$ fixed
and varied $v_s$ from values lower to values higher than $v_w$.
First of all,
we notice that the effusion rates of both JP species are almost insensitive
to the rotational diffusion for $v_s \rightarrow 0$, while develop a strong
dependence on $D_{\theta}$ in the opposite limit, $v_s \rightarrow \infty $.
For $v_s > 10 v_w$, at $D_{\theta} = 1$ the effusion rate is about one order
of magnitude larger  than at $D_{\theta} = 0.1$.

In Fig.~\ref{F6}(b), we examine the consequences of gradually increasing the fraction of guest particles for different values of their self-propulsion speed, $v_s$. 
While all effusion plots exhibit the same general behavior as
in Fig \ref{F6}(a), a few additional features are remarkable:
\newline (i) The effusion rate of the strongly active JPs keeps increasing, but more slowly than $E_m(0)$ in Fig.~\ref{F5}(a), due to their interaction with the less active JPs. 
In such a limit, the most active particles tend to push the less active ones against the box walls. 
Moreover, like in one component systems, clogging effects have great impact on the effusion of both the weak and strong active components. 
\newline (ii) On the contrary, the effusion rate of the weak JPs remains unchanged for $v_s$ up to $v_w$; upon further increasing $v_s$, it goes through a maximum in agreement with the mechanism of effective motility transfer. 
Again, for very large self-propulsion,  $v_s \gg x_L D_\theta, x_L D_\theta$, strongly active JPs jam against the container walls, thus pushing the weaker JPs inside [see snapshots of 
Fig.~\ref{F6}(b) and supplementary FigSM4]. 
Accordingly, the weaker JPs have a small chance to escape through the opening so that effusion becomes drastically suppressed. 
Moreover, no decaying branch of $E_w$ {\it vs.} $v_s$ is detectable at low $\eta_s$. 
This happens because very few strong JPs cannot possibly confine all weak particles in the box interior. 

In conclusion, we stress that adding a small amount of strongly active JPs does suffice to enhance the effusion of sluggish active JPs, but an excess of them can produce the opposite effect!
 Recall that, as illustrated by our simulation snapshots, the two components of an active binary mixture can separate into two distinct phases, when the self-propulsion length of one component is much larger than the size of the container and the other one much shorter, that is, for $v_s /D_\theta \gg x_L,y_L \gg v_w /D_\theta$. 
However, phase segregation should be avoided for better motility transfer. 

As shown in figure \ref{F5}(b), there is a window of the tunable $v_s$, where
the effusion rate of the $w$ particles is enhanced by 2 to 7 times,
depending on their rotational relaxation time and the composition of the binary
mixture. Also, the span of this window is sensitive to the persistence length
of self-propelled motion. This striking result confirms that, even in the
absence of inertia, the {\it motility of the more active microswimmers can be
effectively transferred to the less active microswimmers}.

In our numerical analysis we assumed the pore to be centered in one side of a
square-shaped simulation box. However, sliding boundary conditions as the JPs
move against the cavity walls, can affect their average effusion rate. Our
simulation shows that this may become an issue only at zero temperature. As a
matter of fact, thermal fluctuations assist the escape mechanism by enhancing
particle diffusion along the boundaries, thus suppressing possible effects
related to the cavity geometry and the actual pore location. To verify this
point, we simulated the effusion rate (not shown) for a modified box
geometry, whereby the escape pore was moved toward one corner; for the
simulation parameters of Fig.~\ref{F5} we detected no appreciable variations
of the relevant effusion rates.

\section{Conclusions}\label{Conclusions}

We have analyzed the effects of active nano/micromotors with tunable high
motility in a suspension of particles whose motility cannot be directly
controlled. We showed that by injecting a small fraction of more active Janus
particles one can substantially enhance the motility of other less active
species. Such a motility enhancement was demonstrated for two typical cases:
particles with weak inertia, by studying the velocity distributions for both
species, and for overdamped particles, by comparing their effusion rates.

Our numerical study proves that in dense binary mixtures of active particles,
the width of the velocity distribution of the less active particles linearly
grows with the fraction of more active particles. Thus, the number of
particles moving with larger velocity is considerably enhanced. Moreover, for
an appropriate choice of the mixture parameters, in the overdamped regime the
motility transfer from the more active to the less active subsystem can raise
the effusion rate of the latter by 2 to 7 times.

Such a technique of motility control can be implemented in a large variety of
biological and medical situations, where one wishes to enhance the motility
of insufficiently active nano- or micro-particles. For example, in the case
of weakly motile sperm cells, our proposal has advantages over other similar
proposals (e.g., using self-propelled metallic rotors trapping sperm cells
\cite{sperm}), whereby it is substantially less damaging to living swimmers
and much easier to implement, as it does not require the fast guest swimmers
to localize and trap individual host particles one by one. Another suggestive
application of this method of motility transfer is to speed up a chemical
reaction involving slowly diffusing nano-particles, by adding a small amount
of more active neutral particles as stirrers \cite{sen-enzyme}.

\section*{Conflicts of interest}
There are no conflicts to declare.

\section*{Accknowledgement}
We thank RIKEN Hokusai for providing computational resources.
P.K.G. is supported by SERB Start-up Research Grant (Young Scientist) No. YSS/2014/000853 and UGC-BSR Start-Up Grant No. F.30-92/2015.
D.D. thanks CSIR, New Delhi, India, for support through a Junior Research Fellowship.
V.R.M. and F.N. acknowledge support by the Research Foundation-Flanders (FWO-Vl) and Japan Society for the Promotion of Science (JSPS) (JSPS-FWO Grant No. VS.059.18N). F.N. is supported in part by the:
MURI Center for Dynamic Magneto-Optics via the
Air Force Office of Scientific Research (AFOSR) (FA9550-14-1-0040),
Army Research Office (ARO) (Grant No. Grant No. W911NF-18-1-0358),
Japan Science and Technology Agency (JST) (via the Q-LEAP program, and the CREST Grant No. JPMJCR1676),
Japan Society for the Promotion of Science (JSPS) (JSPS-RFBR Grant No. 17-52-50023, and
JSPS-FWO Grant No. VS.059.18N), the RIKEN-AIST Challenge Research Fund,
the Foundational Questions Institute (FQXi), and the NTT PHI Laboratory.

%%%END OF MAIN TEXT%%%


\balance


%\end{references}
%%%REFERENCES%%%
\bibliography{rsc} %You need to replace "rsc" on this line with the name of your .bib file
\bibliographystyle{rsc} %the RSC's .bst file

\end{document}

\pagestyle{fancy}
\thispagestyle{plain}
\fancypagestyle{plain}{

%%%HEADER%%%
%\fancyhead[C]{\includegraphics[width=18.5cm]{head_foot/header_bar}}
%\fancyhead[L]{\hspace{0cm}\vspace{1.5cm}\includegraphics[height=30pt]{head_foot/journal_name}}
%\fancyhead[R]{\hspace{0cm}\vspace{1.7cm}\includegraphics[height=55pt]{head_foot/RSC_LOGO_CMYK}}
\renewcommand{\headrulewidth}{0pt}
}
%%%END OF HEADER%%%

%%%PAGE SETUP - Please do not change any commands within this section%%%
\makeFNbottom
\makeatletter
\renewcommand\LARGE{\@setfontsize\LARGE{15pt}{17}}
\renewcommand\Large{\@setfontsize\Large{12pt}{14}}
\renewcommand\large{\@setfontsize\large{10pt}{12}}
\renewcommand\footnotesize{\@setfontsize\footnotesize{7pt}{10}}
\makeatother

\renewcommand{\thefootnote}{\fnsymbol{footnote}}
\renewcommand\footnoterule{\vspace*{1pt}% 
\color{cream}\hrule width 3.5in height 0.4pt \color{black}\vspace*{5pt}} 
\setcounter{secnumdepth}{5}

\makeatletter 
\renewcommand\@biblabel[1]{#1}            
\renewcommand\@makefntext[1]% 
{\noindent\makebox[0pt][r]{\@thefnmark\,}#1}
\makeatother 
\renewcommand{\figurename}{\small{Fig.}~}
\sectionfont{\sffamily\Large}
\subsectionfont{\normalsize}
\subsubsectionfont{\bf}
\setstretch{1.125} %In particular, please do not alter this line.
\setlength{\skip\footins}{0.8cm}
\setlength{\footnotesep}{0.25cm}
\setlength{\jot}{10pt}
\titlespacing*{\section}{0pt}{4pt}{4pt}
\titlespacing*{\subsection}{0pt}{15pt}{1pt}
%%%END OF PAGE SETUP%%%

%%%FOOTER%%%
\fancyfoot{}
%\fancyfoot[LO,RE]{\vspace{-7.1pt}\includegraphics[height=9pt]{head_foot/LF}}
%\fancyfoot[CO]{\vspace{-7.1pt}\hspace{13.2cm}\includegraphics{head_foot/RF}}
%\fancyfoot[CE]{\vspace{-7.2pt}\hspace{-14.2cm}\includegraphics{head_foot/RF}}
\fancyfoot[RO]{\footnotesize{\sffamily{1--\pageref{LastPage} ~\textbar  \hspace{2pt}\thepage}}}
\fancyfoot[LE]{\footnotesize{\sffamily{\thepage~\textbar\hspace{3.45cm} 1--\pageref{LastPage}}}}
\fancyhead{}
\renewcommand{\headrulewidth}{0pt} 
\renewcommand{\footrulewidth}{0pt}
\setlength{\arrayrulewidth}{1pt}
\setlength{\columnsep}{6.5mm}
\setlength\bibsep{1pt}
%%%END OF FOOTER%%%

%%%FIGURE SETUP - please do not change any commands within this section%%%
\makeatletter 
\newlength{\figrulesep} 
\setlength{\figrulesep}{0.5\textfloatsep} 

\newcommand{\topfigrule}{\vspace*{-1pt}% 
\noindent{\color{cream}\rule[-\figrulesep]{\columnwidth}{1.5pt}} }

\newcommand{\botfigrule}{\vspace*{-2pt}% 
\noindent{\color{cream}\rule[\figrulesep]{\columnwidth}{1.5pt}} }

\newcommand{\dblfigrule}{\vspace*{-1pt}% 
\noindent{\color{cream}\rule[-\figrulesep]{\textwidth}{1.5pt}} }

\makeatother
%%%END OF FIGURE SETUP%%%

%%%TITLE, AUTHORS AND ABSTRACT%%%
\twocolumn[
  \begin{@twocolumnfalse}
%\vspace{3cm}
\sffamily
\begin{tabular}{m{4.5cm} p{13.5cm} }

\includegraphics{head_foot} & \noindent\LARGE{\textbf{Supplementary figures }} 
\\%Article title goes here instead of the text "This is the title"
%\vspace{0.3cm} & \vspace{0.3cm} \\

%& \noindent\large{Tanwi Debnath,\textit{$^{a}$} and Pulak Kumar Ghosh\textit{$^{b\dag}$}} \\%Author names go here instead of "Full name", etc.

%& \noindent\large{Debajyoti Debnath\textit{$^{1}$}, Pulak Kumar Ghosh\textit{$^{ 1}$}, Vyacheslav R. Misko\textit{$^{ 2,3}$}, Yunyun Li\textit{$^{4}$}, Fabio Marchesoni\textit{$^{4}$}, Franco Nori\textit{$^{2,6}$}} \\%Author names go here instead of "Full name", etc.

%\author{Debajyoti Debnath$^{1}$, Pulak K. Ghosh$^{1}$,  Vyacheslav R. Misko$^Vyacheslav R. Misko{2,3}$, Yunyun Li$^{4}$, Fabio Marchesoni$^{4,%5}$, and Franco Nori$^{2,6}$}

\end{tabular}

 \end{@twocolumnfalse} \vspace{0.6cm}

  ]
%%%END OF TITLE, AUTHORS AND ABSTRACT%%%

%%%FONT SETUP - please do not change any commands within this section
\renewcommand*\rmdefault{bch}\normalfont\upshape
\rmfamily
\section*{}
\vspace{-1cm}

\begin{figure}  \centering
\includegraphics[height=0.315\textwidth,width=0.35\textwidth]{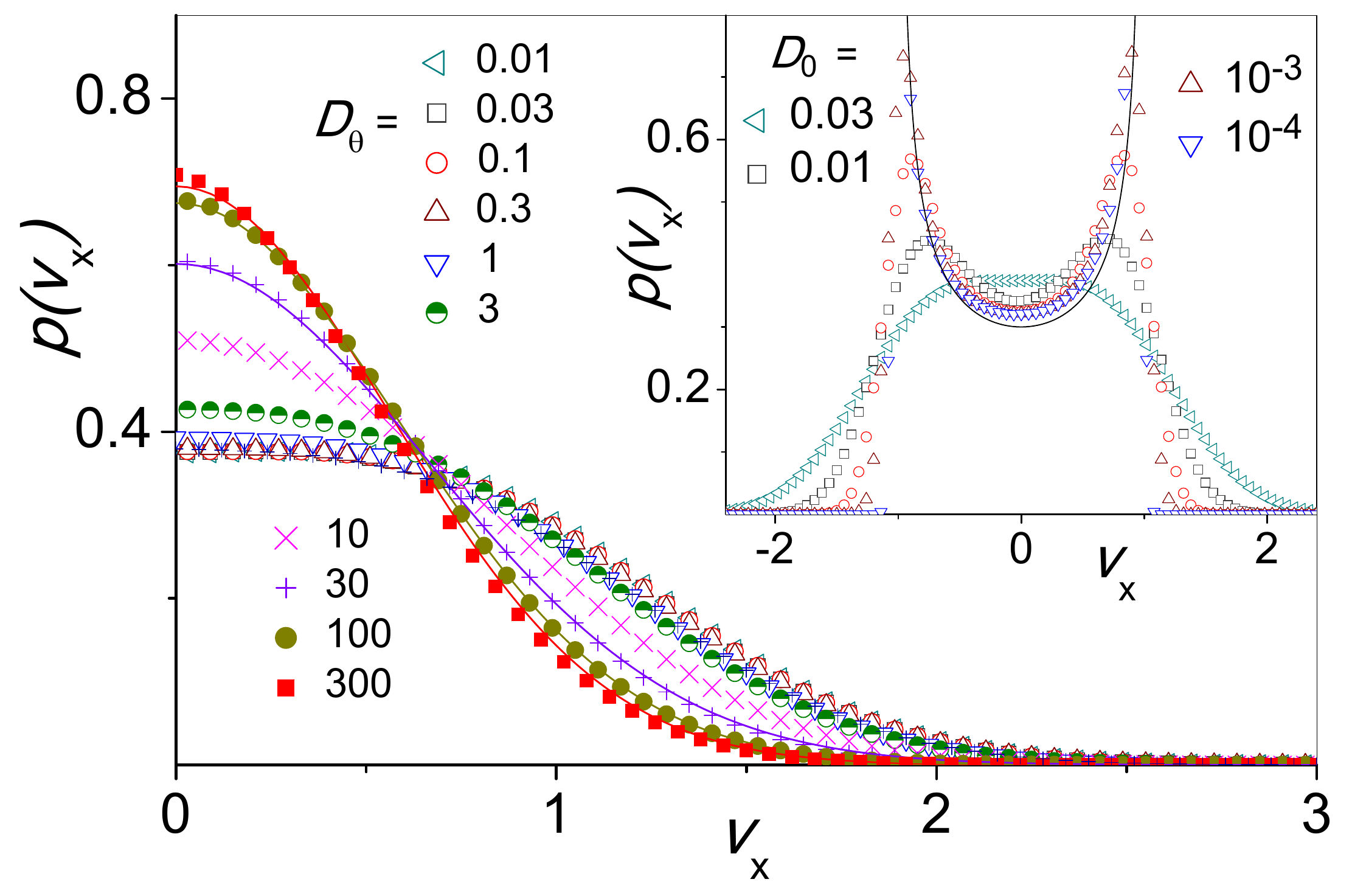}
\nonumber 
\caption* {FigSM1:  (Color online)  Distribution of the $x$-component, $v_x$, of the
velocity for different $D_\theta$ (main panel) and for $D_{\theta} = 0.3$ and
different $D_0$ (inset). Solid lines represent: in the main panel, the
corresponding one-dimensional Maxwell velocity distributions, $p(v_x) =
\sqrt{m/2\pi k T_{\rm eff}}\exp{\left( -mv_x^2/2k T_{\rm eff}\right)}$; in the
inset, the limit for $D_0 \rightarrow 0$, $p(v_x) = 1/\pi \sqrt{1 -
(v_x/v_0)^2}$ .  The parameters used are (unless mentioned in the legends):  $D_0 = 0.03,\; v_0 = 1, \; \gamma =
10, \; m =1 $.  \label {SM1}}
\end{figure}

\begin{figure} [h] \centering
\includegraphics[height=0.315\textwidth,width=0.35\textwidth]{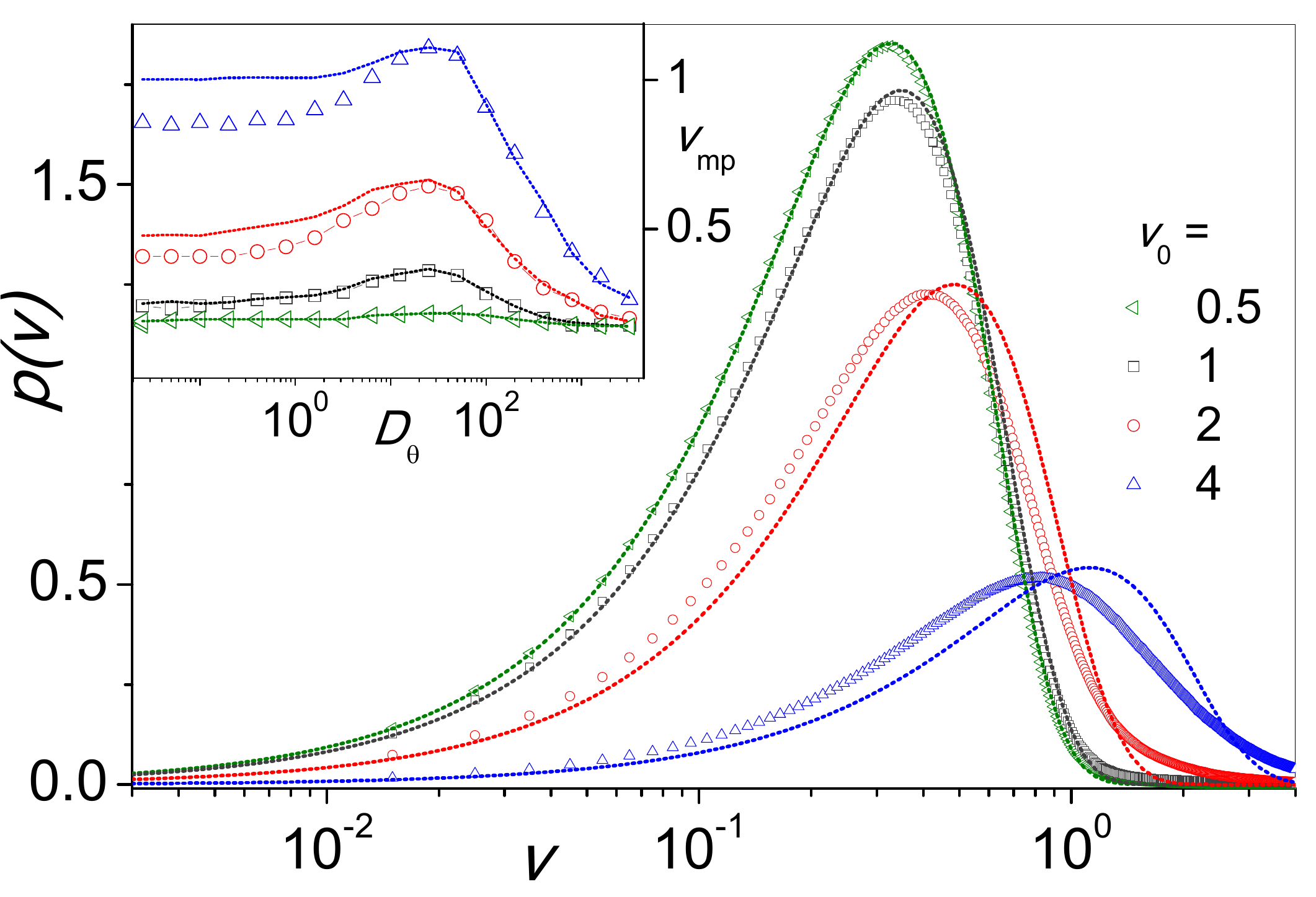}
\nonumber 
\caption* {FigSM2:(Color online) Velocity distribution of interacting self-propelled
particles for different self-propulsion strength $v_0$ . In the main figures,  solid lines are least-square fitting with 2D Gaussian distribution Eq.(8). The parameters used are (unless mentioned in the legends):
 $v_0 = 1, \; \tau_{\theta} = 3.33, \tau_{\gamma} = 0.1, r_0 = 0.75, \; D_0 = 0.01,\; \epsilon = 1,\; \phi = 0.7$. Insets:  $v_{\rm mp}$ versus $D_{\theta}$  for different $v_0$ comparing numerical data with estimates based on $v_{\rm mp} = \sqrt{B}$ (represented by dotted lines).}
\end{figure}

%%%%%%%%%%%%%  Effusion Figures %%%%%%%%%%%%%%%%%%%%%%%%%%%%

\begin{figure} [h] \centering
\includegraphics[height=0.5\textwidth,width=0.4\textwidth]{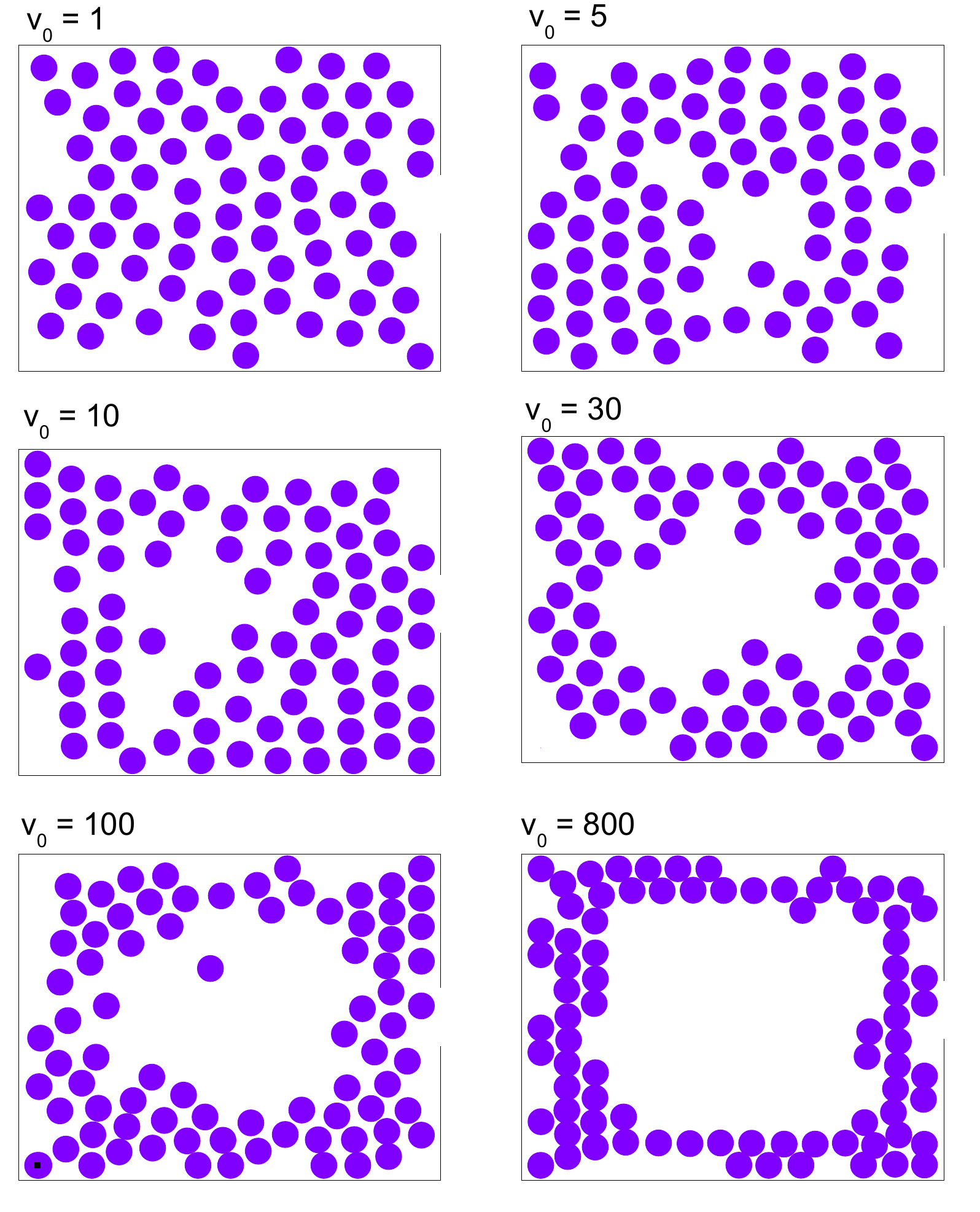}
\nonumber 
\caption* {{\bf FigSM3}: (Color online)  Snapshot of phase points of interacting JPs for different self-propelled velocity $v_0$ and other parameters   $ \epsilon = 1, \; x_{L} = y_{L}=10, \; \Delta = 0.5, r_0 = 0.5,\; D_0 = 0.03,\; N_{t} = 80$,  \label {SE0}}
\end{figure}

\begin{figure} [h] \centering
\includegraphics[height=0.4\textwidth,width=0.4\textwidth]{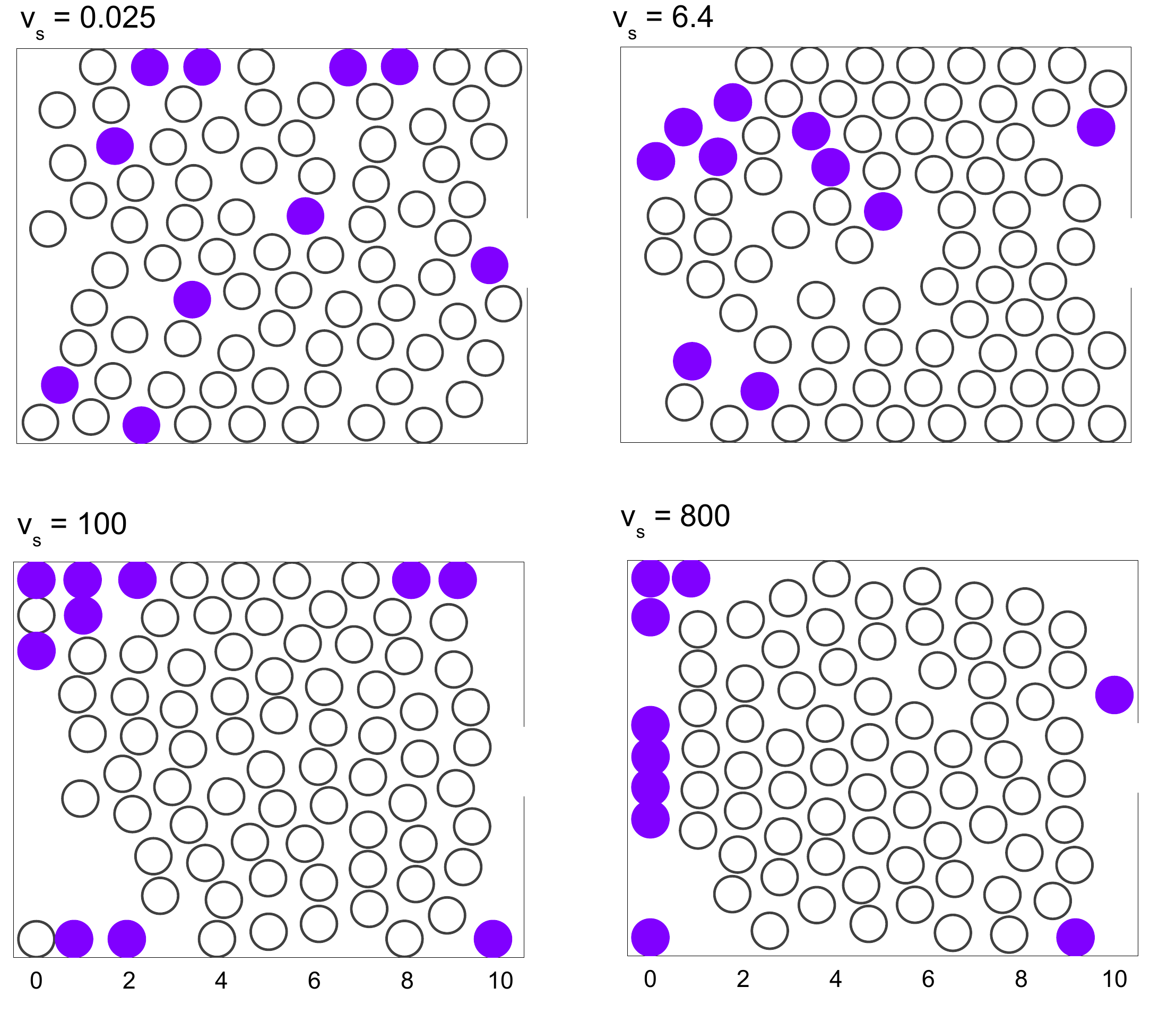}
\nonumber 
\caption* {{\bf FigSM4}: (Color online)  Snapshot of phase points of binary mixture with mole fraction of strong active particles $\eta_s = 0.125$  for different  $v_s$. Filled and empty  circles represent strong and weak JPs, respectively.  Other parameters   $ \epsilon = 1, \; x_{L} = y_{L}=10, \; \Delta = 0.5, r_0 = 0.5,\; D_0 = 0.03,\; N_{t} = 80, \; D_\theta = 1$,   \label {SE0}}
\end{figure}

\begin{figure} [h] \centering
\includegraphics[height=0.4\textwidth,width=0.4\textwidth]{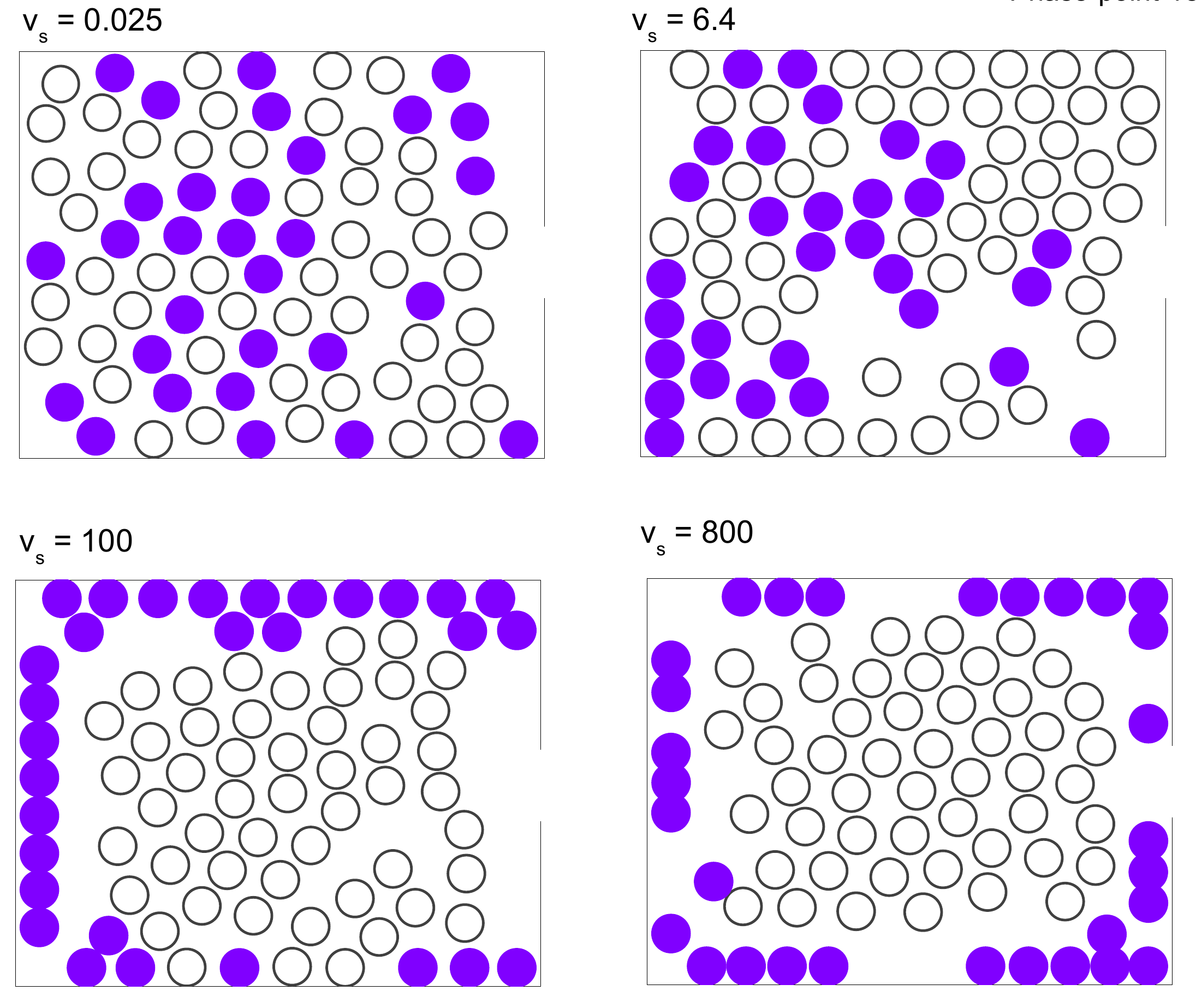}
\nonumber 
\caption* {{\bf FigSM5}: (Color online)  Snapshot of phase points of binary mixture with mole fraction of strong active particles $\eta = 0.375$  for different  $v_s$. Filled and empty  circles represent strong and weak JPs, respectively. Other parameters   $ \epsilon = 1, \; x_{L} = y_{L}=10, \; \Delta = 0.5, r_0 = 0.5,\; D_0 = 0.03,\; N_{t} = 80, \; D_\theta = 1$,   \label {SE0}}
\end{figure}

\begin{figure} [h] \centering
\includegraphics[height=0.4\textwidth,width=0.4\textwidth]{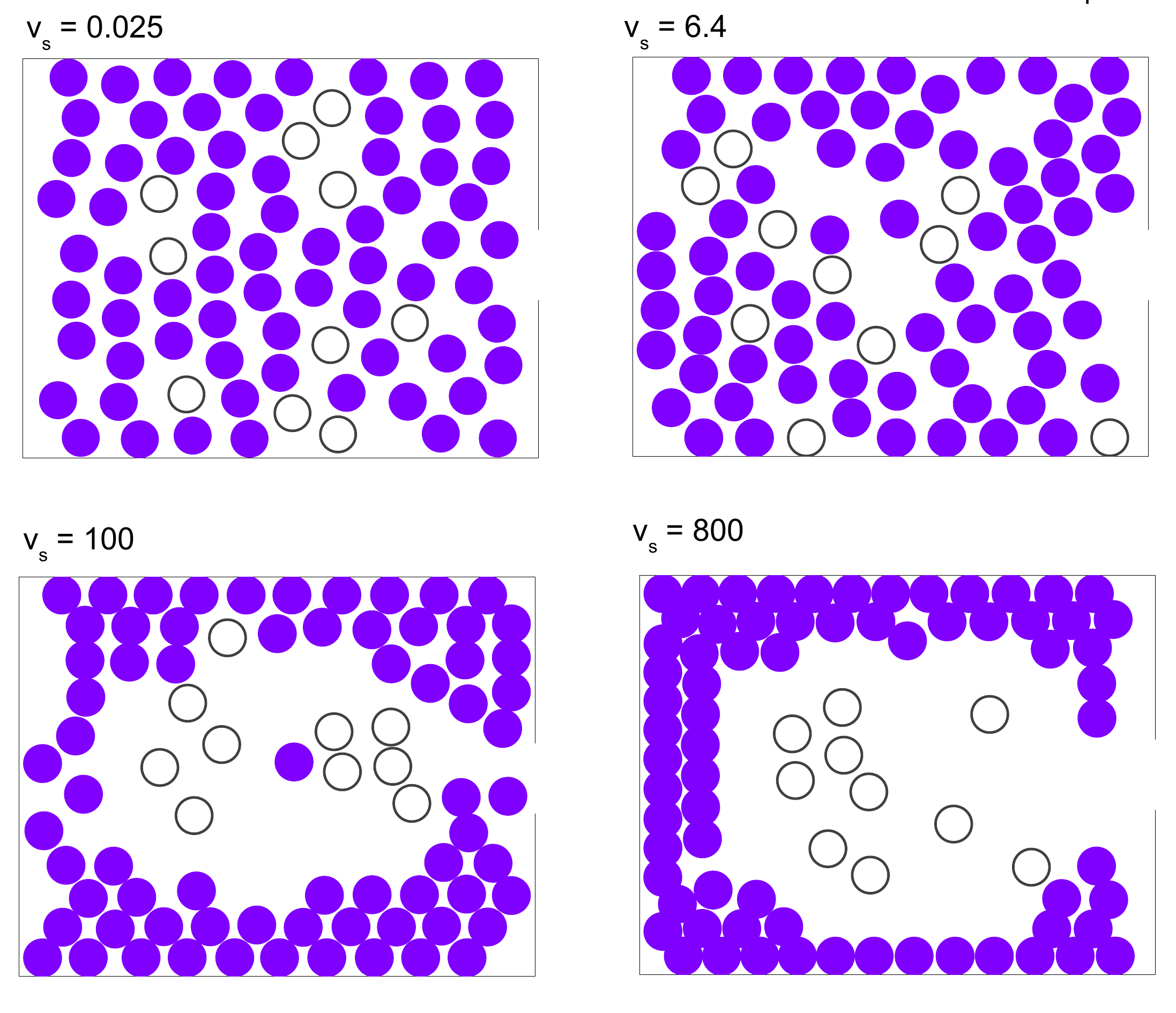}
\nonumber 
\caption* {{\bf FigSM6}: (Color online)  Snapshot of phase points of binary mixture with mole fraction of strong active particles $\eta = 0.88$  for different  $v_s$.  Filled and empty  circles represent strong and weak JPs, respectively. Other parameters   $ \epsilon = 1, \; x_{L} = y_{L}=10, \; \Delta = 0.5, r_0 = 0.5,\; D_0 = 0.03,\; N_{t} = 80, \; D_\theta = 1$,   \label {SE0}}
\end{figure}